\documentclass[iop, numberedappendix]{emulateapj}

\usepackage{apjfonts}

\usepackage[hyperfootnotes=false,naturalnames=true,letterpaper,pdfstartview=FitH,pdfpagemode=UseNone]{hyperref}
\gdef\UmV{U-V}
\gdef\VmJ{V-J}
\gdef\SFRuvir{SFR_\mathrm{UV+IR}}
\gdef\peryr{\mathrm{yr}^{-1}}

\gdef\24mum{$24\,\mu\mathrm{m}$}

\gdef\uJy{\mu\mathrm{Jy}}
\gdef\LIR{L_\mathrm{IR}}

\gdef\4ang{4000\,\AA}

\gdef\mathmsun{\ M_{\odot}}

\gdef\pegase{\textsc{P\'{e}gase}}
\gdef\eazy{\textsc{Eazy}}
\gdef\interrest{\textsc{InterRest}}
\gdef\fast{\textsc{fast}}
\gdef\zCOSMOS{$z$\textsc{COSMOS}}

\gdef\zphot{z_\mathrm{phot}}

\gdef\micron{\mbox{$\mu\mathrm{m}$}}
\newcommand{\weblink}[1]{\href{#1}{#1}}

\shortauthors{Brammer et al.}
\shorttitle{Galaxy number and mass densities to $z=2.2$}
\slugcomment{Submitted to the Astrophysical Journal}
\begin{document}

\title{The number density and mass density of star-forming and quiescent
galaxies at $0.4\leq z\leq 2.2$}

\author{Gabriel~B.~Brammer\altaffilmark{1, 2, 10}, 
        K.~E.~Whitaker\altaffilmark{2}, 
        P.~G.~van~Dokkum\altaffilmark{2}, 
        D.~Marchesini\altaffilmark{3}, 
        M.~Franx\altaffilmark{4},
        M.~Kriek\altaffilmark{5,6}, 
        I.~Labb\'e\altaffilmark{7},
		K.-S.~Lee\altaffilmark{2},
		A.~Muzzin\altaffilmark{2},
		R.~F.~Quadri\altaffilmark{8},
		G.~Rudnick\altaffilmark{9},
		R.~Williams\altaffilmark{8}
}

\email{gbrammer@eso.org}

\altaffiltext{1}{European Southern Observatory, Alonso de C\'ordova 3107, Casilla 19001, Vitacura, Santiago, Chile}

\altaffiltext{2}{Department of Astronomy, Yale University, New Haven, CT 06520.}
\altaffiltext{3}{Department of Physics and Astronomy, Tufts University, Medford, MA 02155}
\altaffiltext{4}{Leiden Observatory, P.O. Box 9513, NL-2300 RA, Leiden, Netherlands.}
\altaffiltext{5}{Department of Astrophysical Sciences, Princeton University,
Princeton, NJ 08544.}
\altaffiltext{6}{Harvard-Smithsonian Center for Astrophysics, 60 Garden Street, Cambridge, MA 02138}
\altaffiltext{7}{Carnegie Observatories, 813 Santa Barbara Street, Pasadena, CA 91101.}
\altaffiltext{8}{Carnegie Observatories, 813 Santa Barbara Street, Pasadena, CA 91101.}
\altaffiltext{9}{The University of Kansas, Department of Physics 
and Astronomy, Malott room 1082, 1251 Wescoe Hall Drive, 
Lawrence, KS, 66045.}
\altaffiltext{10}{Visiting Astronomer, Kitt Peak National Observatory, National Optical Astronomy Observatory, which is operated by the Association of Universities for Research in Astronomy (AURA) under cooperative agreement with the National Science Foundation.}

\begin{abstract}
    
We study the build-up of the bimodal galaxy population using the NEWFIRM
Medium-Band Survey, which provides excellent redshifts and well-sampled spectral
energy distributions of $\approx 27,000$ galaxies with $K<22.8$ at $0.4<z<2.2$.
We first show that star-forming galaxies and quiescent galaxies can be robustly
separated with a two-color criterion over this entire redshift range. We then
study the evolution of the number density and mass density of quiescent and
star-forming galaxies, extending the results of the COMBO-17, DEEP2, and other
surveys to $z=2.2$. The mass density of quiescent galaxies with $M\gtrsim
3\times 10^{10}$\,$M_{\odot}$ increases by a factor of $\sim 10$ from $z\sim 2$
to the present day, whereas the mass density in star-forming galaxies is flat or
decreases over the same time period. Modest mass growth by a factor of $\sim 2$
of individual quiescent galaxies can explain roughly half of the strong density
evolution at masses $>10^{11}$\,$M_{\odot}$, due to the steepness of the
exponential tail of the mass function. The rest of the density evolution of
massive, quiescent galaxies is likely due to transformation (e.g. quenching) of
the massive star-forming population, a conclusion which is consistent with the
density evolution we observe for the star-forming galaxies themselves, which is
flat or decreasing with cosmic time. Modest mass growth does not explain the
evolution of less massive quiescent galaxies ($\sim 10^{10.5}~M_\odot$), which
show a similarly steep increase in their number densities. The less massive
quiescent galaxies are therefore continuously formed by transforming galaxies
from the star-forming population.

\end{abstract}


%%%%%%%%%%%%%%%%%%%%%%%%%%%%%%%%%%%%%%%%%%%%%%%%%%%%%%%%%%%%%%%%%%
%
%   Introduction
%
%%%%%%%%%%%%%%%%%%%%%%%%%%%%%%%%%%%%%%%%%%%%%%%%%%%%%%%%%%%%%%%%%%
\section{Introduction}\label{s:intro}

Large surveys, such as the Sloan Digital Sky Survey (SDSS),
have begun to sample representative volumes of the
nearby Universe.
One of the more surprising results from these
surveys is the
existence of a bimodal galaxy population, manifested in correlations
between a wide variety of galaxy properties both observed
(i.e., color vs. luminosity, color vs. morphology;
\citealp{strateva:01, blanton:03, baldry:04}) and derived (i.e.,
stellar age and stellar mass; \citealp[e.g.][]{kauffmann:03}).  This
bimodal population is composed of red, early-type galaxies with old
stellar populations and little ongoing star-formation that tend to be
the most luminous and massive galaxies at any redshift, and a
complementary population of star-forming disk galaxies with bluer
colors typical of young stellar populations.

A key question is when this bimodality was established, and what
fraction of the total stellar mass is locked
up in each of the two galaxy types as a function of cosmic time.
In a landmark study,
\cite{bell:04} find that a red sequence was already in place
at $z\sim1$.
The color evolution of the red sequence at $z<1$ is roughly consistent
with passive evolution, but analyses of the luminosity function
indicate a buildup of a factor of $\sim$2 in stellar mass over
this redshift range
\citep{bell:04, borch:06, arnouts:07, brown:07, faber:07,
  ilbert:10}. The results at $z\sim1$ are somewhat
uncertain because they require large corrections for incompleteness
\citep[see][]{faber:07}.  Furthermore, the most massive galaxies do
not appear to evolve significantly at $0<z<1$ \citep{wake:06, brown:07},
implying that they were assembled at higher redshifts.

A number of recent studies have extended this work 
to $z\sim2$. In particular,
\cite{arnouts:07} and \cite{ilbert:10} find an increase of a factor of
$\sim$10 in the stellar mass density of quiescent galaxies between
$z=2$ and $z=1.2$, which would imply very dramatic changes over a 
relatively short ($\approx 2$\,Gyr) period.
These studies are not definitive, as they depend on relatively uncertain
photometric redshifts at $z>1$ determined from broad-band NIR
photometry. \cite{taylor:09b} show that large redshift uncertainties
make the robust identification of a bimodal galaxy population
extremely difficult at $z>1.5$. Unfortunately, spectroscopic verification
of these results is extremely difficult due to the faintness of massive
galaxies in the observer's optical \citep[see, e.g.,][]{kriek:08}.

Here we examine the evolution of the bimodal galaxy population with
the NEWFIRM Medium-Band Survey (NMBS), a moderately deep, moderately
wide near-IR survey which uses a novel set of medium-bandwidth filters
specifically tuned to the redshift range $1<z<3$ \citep{nmbs, whitaker:10}.
In \cite{brammer:09}, we used the NMBS to show that
massive galaxies are nearly all red up to $z\sim2$ but that a bimodal
population is apparent after correcting the colors of galaxies heavily
reddened by dust.  \cite{whitaker:10} find that not only can quiescent
galaxies be identified in the NMBS up to $z\sim2$, but their color
scatter within the red sequence is resolved.  They find that the
fraction of quiescent galaxies among all galaxies at $M >
10^{11}\mathmsun$ decreases sharply from roughly 90\% at $z=1$ to 40\%
at $z=2$.  Furthermore, \cite{marchesini:10} find evidence that the quiescent fraction of extremely massive galaxies ($M>10^{11.3}M_\odot$) decreases further still to 7--30\% by $z=3.5$.

In this paper, we use the NMBS to
study the buildup of star-forming and quiescent galaxies
from $z=2$ to the present by constructing their mass functions
and quantifying the evolution of their number and mass densities. In
\S\ref{s:data} we describe the survey data and sample selection.  We
show the galaxy rest-frame color distribution and its evolution in
\S\ref{s:color_mass_diagram}, and describe a method of cleanly
separating the red, ``quiescent'' galaxy sequence from intrinsically blue,
star-forming galaxies by accounting for the effects of dust reddening.
We present stellar mass functions in \S\ref{s:mass_functions}. We
study the evolution of the galaxy number and mass densities in
\S\ref{s:density}, and we discuss our results in the context of
massive galaxy formation in \S\ref{s:discussion}.  We summarize our
results in \S\ref{s:summary}.  We assume a $\Lambda$CDM cosmology
throughout, with $\Omega_m=0.3$, $\Omega_\Lambda=0.7$, and
$H_0=70\ \mathrm{km\ s}^{-1}\ \mathrm{Mpc}^{-1}$.  All magnitudes and
colors are given in the AB system.

% \textbf{Goals of the paper:} 
% \begin{enumerate}
% 	\item Number and mass density evolution of red sequence galaxies to $z=2$.
% 	\item How does the disappearance of massive (dusty) star-forming galaxies compare to the buildup of massive quiescent galaxies?  
% \end{enumerate}

%%%%%%%%%%%%%%%%%%%%%%%%%%%%%%%%%%%%%%%%%%%%%%%%%%%%%%%%%%%%%%%%%%
%
%   Data / NMBS description
%
%%%%%%%%%%%%%%%%%%%%%%%%%%%%%%%%%%%%%%%%%%%%%%%%%%%%%%%%%%%%%%%%%%
\section{Data}\label{s:data}

\subsection{The NEWFIRM Medium-Band Survey}\label{s:survey}

The NEWFIRM Medium-Band Survey \citep[NMBS;][]{nmbs, whitaker:11} provides
well-sampled galaxy spectral energy distributions (SEDs) from
rest-frame UV through NIR wavelengths up to $z\sim 3.5$, thanks in
part to a custom set of five medium-band NIR filters that span
observed wavelengths 1--1.7~\micron\ at roughly twice the spectral
resolution of standard broad-band filters.  The medium-band filter
technique has been successfully employed at optical wavelengths to
measure very precise photometric redshifts at $z\lesssim1.4$
(COMBO-17, \citealp{wolf:03}; COSMOS, \citealp{ilbert:09}; E-CDFS,
\citealp{cardamone:10}).  The ability of medium-band filters to
constrain photometric redshifts depends on their sampling strong,
broad spectral features, and in particular the
Balmer/4000\,\AA\ break, which is redshifted to
$\lambda>1~\micron$\ at $z\gtrsim1.5$.  The NMBS filters are designed
to enable precise photometric redshift estimates at $1.5 < z < 3.5$ by
improved sampling of the Balmer/4000\,\AA\ break at these redshifts.
We briefly summarize the NMBS below; a full description of the data
reduction and photometric catalogs is provided by \cite{whitaker:11}.

The NMBS provides medium-band near-IR photometry over $\sim$0.25 deg$^2$ NEWFIRM \citep{probst:newfirm} pointings in each of two well-studied survey fields, COSMOS \citep{scoville:cosmos} and the All-wavelength Extended Groth Strip International Survey (AEGIS)\footnote{\weblink{http://aegis.ucolick.org/}}.  The NEWFIRM data were taken over 75 nights in 2008--9 at the Mayall/4m telescope at the Kitt Peak National Observatory.  The NEWFIRM $J123$, $H12$, and $K$ data are supplemented at optimal wavelengths by deep $ugriz$ imaging in both fields from the CFHT Legacy Survey\footnote{\weblink{http://cfht.hawaii.edu/Science/CFHTLS}}, as reduced by the CARS team \citep{erben:09, hildebrandt:09}.  Furthermore, we include deep Subaru imaging in broad-band $B_JV_Jr^+i^+z^+$ and in 12 medium-band filters that span 4000--8000\,\AA\ \citep{capak:07, ilbert:09}.  At mid-IR wavelengths, we include \textit{Spitzer}-IRAC 3--8~\micron\ and MIPS 24~\micron\ imaging that cover the entire COSMOS pointing \citep[S-COSMOS;][]{sanders:07} and 60\% of the AEGIS pointing \citep{barmby:08}.  After masking regions with less than 30\% of the maximum exposure time in the NMBS bands and regions around bright stars, the NMBS covers \mbox{0.20 deg$^2$} and \mbox{0.19 deg$^2$} in COSMOS and AEGIS, respectively.

Objects are detected in the NEWFIRM $K$ image, and the optical/NIR
photometry is performed on images convolved to the same point-spread
function (PSF) to limit band-dependent effects (\citealp{whitaker:11}; see \citealt{quadri:07} for a similar photometric strategy).
Objects $K=22.8$ are detected at $\sim$5$\sigma$, and
the corresponding depths in the medium-bands are approximately flat as
a function of $f_\lambda$.  The CFHT and Subaru broad-band optical
images are among the deepest available in any field, which is
especially important for producing high-S/N SEDs of optically-faint,
red galaxies at $z>1$.  The large-PSF IRAC and MIPS images require
advanced photometric techniques to minimize photometric contamination
by neighboring objects.  We employ a source-fitting method that uses
the higher-resolution $K$-band image to model the positions and sizes
of objects in the redder \textit{Spitzer} bands, whose flux
normalizations are fit by least-squares regression.  Fluxes and errors
are then measured for each object with simple aperture photometry
after subtracting the model of all neighboring objects
(\citealp{labbe:06}; see \citealp{wuyts:07} for an illustrative
example).  The IRAC images are significantly deeper than the NEWFIRM
$K$-band.  The MIPS 24~\micron\ images reach $\sim$20 $\uJy$ at
$3\sigma$.

The primary sample used throughout this paper is defined as all
galaxies in the \cite{whitaker:11} catalog (version 5.0)
of the two NMBS fields with $K<22.8$, $0.4<\zphot<2.2$, and
the standard quality cuts, which yields 25,423
galaxies (15,485 with MIPS coverage).

\subsection{Photometric Redshifts}\label{s:photoz}

We estimate photometric redshifts from the $u\mbox{--}8\,\micron$ SEDs using the \eazy\ photometric redshift code \citep{brammer:08}.  We use the default \eazy\ template set described by \cite{brammer:08} with a modified treatment of emission lines inspired by \cite{ilbert:09}; rather than using the emission lines as predicted by the \pegase\ population synthesis code \citep{pegase}, we compute a ``star-formation rate'' (SFR) from the rest-frame 2800\,\AA\ flux of each template and then add $H\alpha\mbox{--}\gamma$, Lyman-$\alpha$, \ion{O}{2} $\lambda3727$ and \ion{O}{3} $\lambda\lambda4959, 5007$ emission lines using fixed line ratios (\citealp{ilbert:09}; after \citealp{kennicutt:98}).  Although this treatment of emission lines is still over-simplified---real galaxies will have a non-trivial range of SFRs, line strengths and line ratios for a given 2800\,\AA\ flux---we find that it significantly improves the photometric redshift quality of the medium-band SEDs, which are more sensitive to line contamination than broad-band SEDs (see also \citealt{ilbert:09} and Appendix \ref{ap:emlines} below).

To further improve the photometric redshifts, we iteratively adjust the photometric zeropoints of the ground-based photometric bands to minimize the residuals to template fits at fixed redshift for objects with spectroscopic redshifts (see below).  With the exception of CFHT-$u$, which appears to be $\sim$0.2 mag ``too faint'' (\citealt{erben:09} note a similar discrepancy), the adjustments to the CFHT/NEWFIRM zeropoints are $<$2\%, well within typical $\sim$0.05 mag zeropoint calibration uncertainties.  The Subaru broad- and medium-band images require significantly larger corrections to their publicly-listed zeropoints of up to 0.2 mag \citep[see also Table 1 of ][]{ilbert:09}.  These offsets are constrained by the well-calibrated, overlapping CFHT photometry, and we find that the photometric redshift quality is improved significantly when including all of the available optical data.    

Two large spectroscopic surveys, DEEP2 \citep{davis:deep2} and \zCOSMOS\ \citep{lilly:07}, provide a large number of spectroscopic redshift measurements that can help us assess the quality of our photometric redshift estimates.  For 2067 objects from our sample that have DEEP2 redshifts, we measure a NMAD \citep[see][]{brammer:08} scatter $\sigma/(1+z)=0.016$.  In the COSMOS field, where objects have 35-band SEDs that include medium bands from 4000\,\AA--1.7\micron, the NMAD scatter is $\sigma/(1+z)=0.008$ for 1099 galaxies with \zCOSMOS\  redshifts.\footnote{We use only the most reliable spectroscopic redshift quality flags from DEEP2 ($4$) and \zCOSMOS\ ($3.x, 4.x$).}  The number of catastrophic redshift failures with $|\Delta z/(1+z)| > 0.1$ is 2.6\% (1.6\%) in AEGIS (COSMOS).  These results are among the best that have been achieved for photometric surveys, and we note that \eazy\ has been shown to produce redshift estimates with the lowest scatter and the smallest fraction of outliers of any of the public photometric redshift codes \citep{hildebrandt:10}.

We caution that the subset of objects with spectroscopic redshifts tend to be optically-bright (e.g., $i<22.5$ for \zCOSMOS) and almost all are at $z<1.5$; the spectroscopic sample is not representative of the full $K$-selected sample used throughout this paper.  \cite{nmbs} targeted four galaxies in the \cite{kriek:08} GNIRS spectroscopic sample to test the NEWFIRM medium-band technique on $K$-selected galaxies at $z>1.7$, and they find $\Delta z/(1+z)<0.02$ for all four objects.  Furthermore, \cite{kriek:11} have shown recently that H$\alpha$ emission can be detected in composite SEDs built from the medium-band photometry itself, and they conclude that the photometric redshift errors must be $\lesssim 2\%$ given the shape of the observed emission line.  While small spectroscopic samples suggest  the potential of the NMBS medium-band filters for providing accurate photometric redshifts at $z>1.5$, further spectroscopic followup is required to fully assess the redshift quality at the redshift and magnitude limits of the survey.

\subsection{Rest-frame colors}\label{s:rf_colors}

We derive rest-frame $\UmV$ and $V-J$ colors\footnote{We adopt the $U$ and $V$
  filter definitions of \cite{maiz:06}, and 2MASS-$J$} from the template that best
fits the observed photometry, which itself is a linear combination of
the \eazy\ templates.  We take the rest-frame color from the template
directly, similar to the technique described by \cite{wolf:03} for the
COMBO-17 survey.\footnote{\eazy2.0 with the updated templates and the
  rest-frame color implementation is made available at
  \weblink{http://www.astro.yale.edu/eazy/}.}  Other methods have been
developed to measure rest-frame fluxes by interpolating between the
observed bands that bracket the rest-frame band at a given redshift,
with the potential advantage that the resulting rest-frame fluxes are
determined more directly from the observed photometry rather than
depending on the choice of fitting templates (e.g., the algorithm of
\citealp{rudnick:03}, implemented in the \interrest\ program by
\citealp{taylor:09b}).  For a sufficiently flexible set of templates
(e.g., arbitrary linear combinations of the \eazy\ templates), these
two methods produce very similar results for photometric surveys with
mostly non-overlapping, broad-band filters.  As the filters become
more closely-spaced, direct template fluxes are more robust as they
are able to make use of more of the observed photometry.  That is, at
a particular redshift the interpolation method may interpolate between
two adjacent filters that have relatively low S/N, while the SED shape
at the desired wavelength is actually well-constrained by additional
nearby filters.  We adopt the direct template $\UmV$ colors
throughout, but note that our results are qualitatively consistent for
both our template-fitting and the \cite{taylor:09b} methods.

%%%%%%%%%%%%%%%%%%%%%%%% Color-mass diagrams   %%%%%%%%%%%%%%%%%%%%
\begin{figure*}
	\epsscale{1.15}
	\plotone{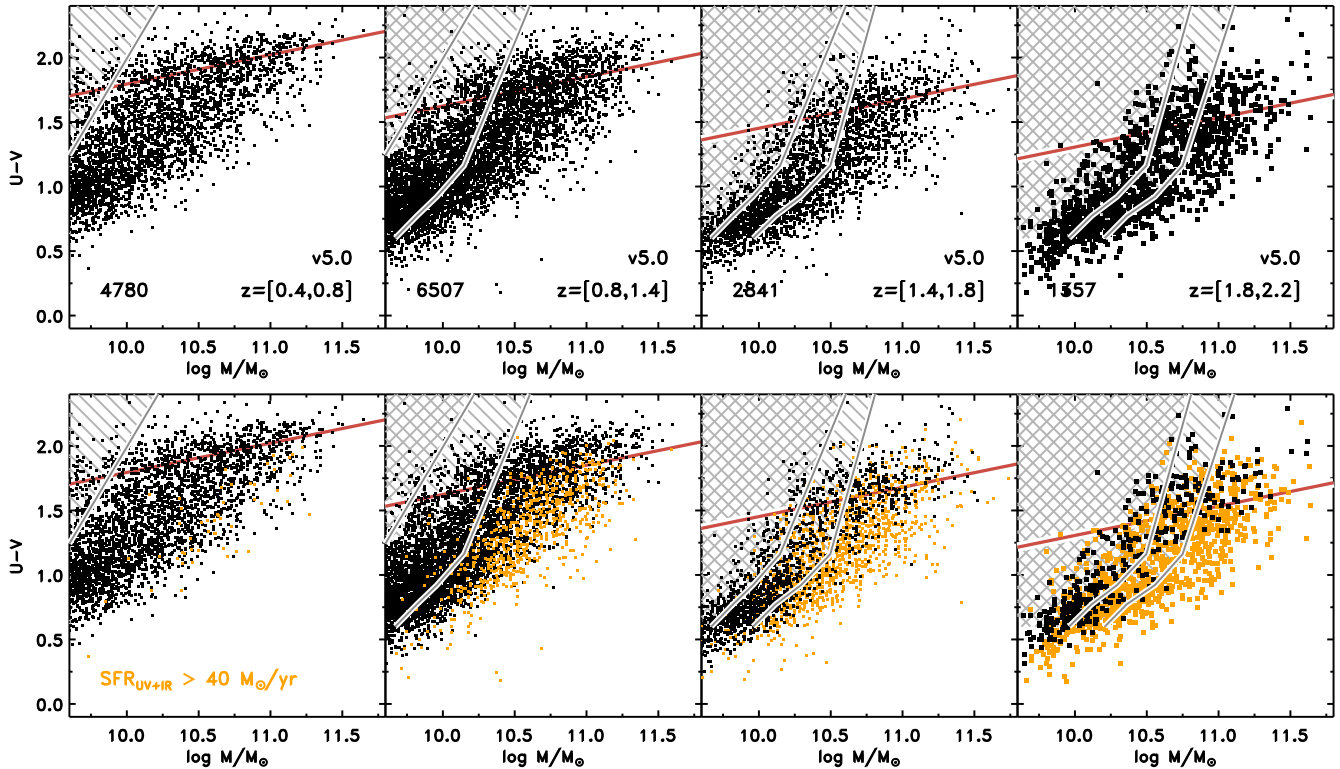}

	\caption[Color-mass diagrams at $0.4 < z < 2.2$]{\textit{Top panels:}  Color-mass diagrams at $0.4 < z < 2.2$. The rest-frame $U-V$ colors are derived
from the SEDs as described in \S\ref{s:rf_colors}. The grey lines and hatched
regions indicate where the NMBS is less than 90\% complete at the edges of the
redshift bins (i.e. the region defined by the left-most of the two lines
corresponds to the low-$z$ edge of the bin). The number of galaxies in each bin
is indicated at lower-left and the symbol size is roughly inversely proportional
to the bin sample size for clarity. The solid red line in the top panels
indicates the red-sequence found by \cite{borch:06} at $z<1$, with the redshift
evolution of the red-sequence color zeropoint extrapolated to $z=2$. Note that
all massive galaxies are red, out to the highest redshifts. {\em Bottom panels:}
Same as the top panels, but galaxies with $\SFRuvir>40\ M_\odot~\peryr$ are
shown in orange. Many of the massive red galaxies at $z\sim 2$ are dusty
star-forming systems, which means that a division of the galaxy population by a
single rest-frame color does not lead to a homogeneous galaxy population.  Note that only galaxies with MIPS coverage, necessary for estimating the SFR, are included in this figure.
\label{f:red_sequence}}

\end{figure*}

\subsection{$\LIR$ and star formation rates}\label{s:sfrs}

We use the rest-frame color routine described above to derive $\nu I_\nu$ luminosities at 2800~\AA, which we combine with the MIPS 24~\micron\ photometry to estimate star formation rates, $\SFRuvir = 0.98\times10^{-10}\left(\LIR+3.3L_{2800}\right)$ (\citealp{bell:05}; adapted for the Kroupa IMF by \citealp{franx:08}).  The MIPS 24\micron\ fluxes are converted to total IR luminosities ($\LIR = L_{8-1000~\micron}$) using the \cite{dale:02} templates, where we adopt the log average conversion for templates with $1 < \alpha < 2.5$ \citep[see][]{wuyts:fireworks}.  This conversion has a systematic uncertainty of a factor of $\sim$3 \citep{wuyts:fireworks}. \cite{papovich:07} and \cite{muzzin:10} (among others) show that the commonly-used alternative method of fitting the luminosity-dependent \cite{chary:01} templates tends to overestimate $\LIR$ at $z\gtrsim1.5$ by factors of $\sim$5.

\subsection{SED modeling: stellar masses, $A_V$}\label{s:fast}

We measure stellar masses and dust reddening of the NMBS galaxies by
fitting a grid of population synthesis models to the NMBS SEDs using
the \fast\ code \citep{kriek:09a}.  The model grid is composed of
\cite{bc:03} models with exponentially-declining star formation
histories (SFHs) with decline rates $\log \tau/\mathrm{yr}=7\mbox{--}10$,
computed with a \cite{kroupa:01} initial mass function (IMF).  To
first order, alternative choices of the IMF
\citep[e.g.][]{salpeter:55} cause a shift in the derived stellar
masses without affecting other properties of the fit (e.g.,
$M_\mathrm{Salp.}/M_\mathrm{Kroupa}\sim1.6$; \citealp{marchesini:09}).  Note that adopting alternative star-formation histories, such as a SFH that increases with time \citep{maraston:10, papovich:11}, could also affect the derived stellar masses by significant factors, perhaps as large as a factor of two.
We estimate the amount of dust reddening from the SED fit by allowing
a uniform dust screen with up to four magnitudes of extinction in the
$V$-band ($A_V=0\mbox{--}4$) and with a wavelength dependence
following the \cite{calzetti:00} reddening law.  Using a similar
SED-fitting technique with photometric redshifts and broad-band
photometry, \cite{muzzin:09} demonstrate that stellar masses
can be determined with precision $\sim$0.1 dex.
\cite{marchesini:09} show, however, that changing the modeling
assumptions such as the modeling library or the IMF can cause
systematic differences in the derived stellar masses that are larger
than these random errors.  For example, stellar masses estimated with
\cite{maraston:05} models tend to be systematically lower than those
determined from \cite{bc:03} models by a factor of $\sim$1.4
\citep{wuyts:07, whitaker:10}. We use the \cite{bc:03} models as
they appear to better describe the SEDs of young quiescent galaxies
than the \citet{maraston:05} models \citep{conroy:09, kriek:10}.

%%%%%%%%%%%%%%%%%%%%%%%%%%%%%%%%%%%%%%%%%%%%%%%%%%%%%%%%%%%%%%%%%%
%
%   Sample definition
%
%%%%%%%%%%%%%%%%%%%%%%%%%%%%%%%%%%%%%%%%%%%%%%%%%%%%%%%%%%%%%%%%%%
\section{The bimodal galaxy population}\label{s:color_mass_diagram}

\subsection{Color-mass relations and completeness}

%%%%%%%%%%%%%%%%%%%%%%%%%%%%%%%%%%%%%%%%%%%%%%%%%%%%%%%%%%%%%%%%%%%%%%

\begin{figure*}
	\epsscale{1.15} \plotone{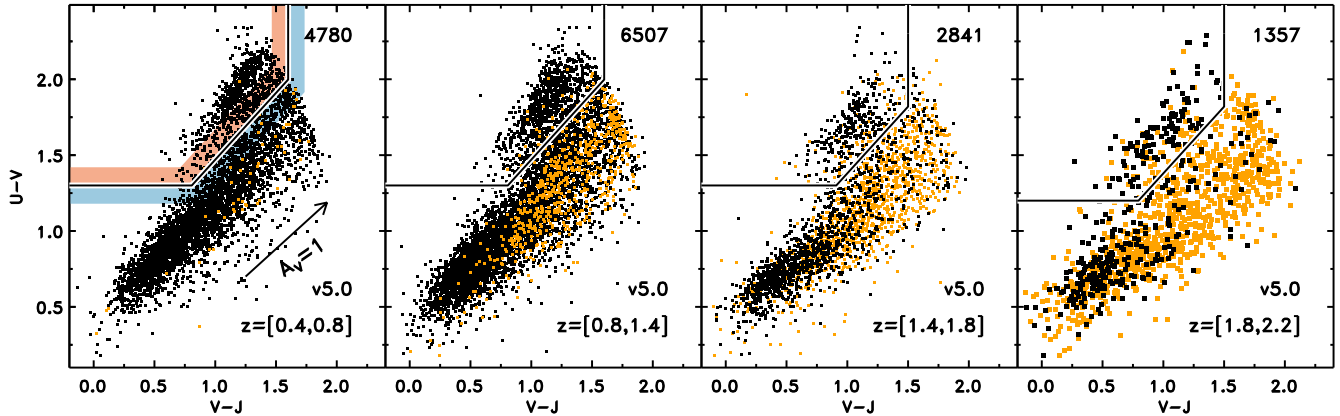}

	\caption[Color-mass diagrams at $0.4 < z < 2.2$]{
Dusty star forming galaxies can be cleanly separated from quiescent galaxies
when using two rest-frame colors. The panels show the galaxy distribution in the
rest-frame $\UmV$ versus $\VmJ$ plane for the same redshift bins as Fig.\ 1.
Galaxies with $\SFRuvir>40\ M_\odot~\peryr$ are again shown in orange. These
galaxies occupy a region that is distinct from the quiescent galaxies. The solid
line and colored bands indicate the red/blue or quiescent/dusty+star-forming
selection developed by \cite{labbe:05} and \cite{williams:09}. The reddening
vector for one magnitude of extinction in the $V$-band is indicated, assuming a
\cite{calzetti:00} reddening law.  Again, only galaxies with MIPS coverage are included in this figure.\label{f:color_color}}

\end{figure*}

%%%%%%%%%%%%%%%%%%%%%%%%%%%%%%%%%%%%%%%%%%%%%%%%%%%%%%%%%%%%%%%%%%%%%%

%%%%%%%%%%%%%%%%%%%%%%%%    Histograms   %%%%%%%%%%%%%%%%%%%%%%%%%    
\begin{figure*}
	\epsscale{0.9}
	\plotone{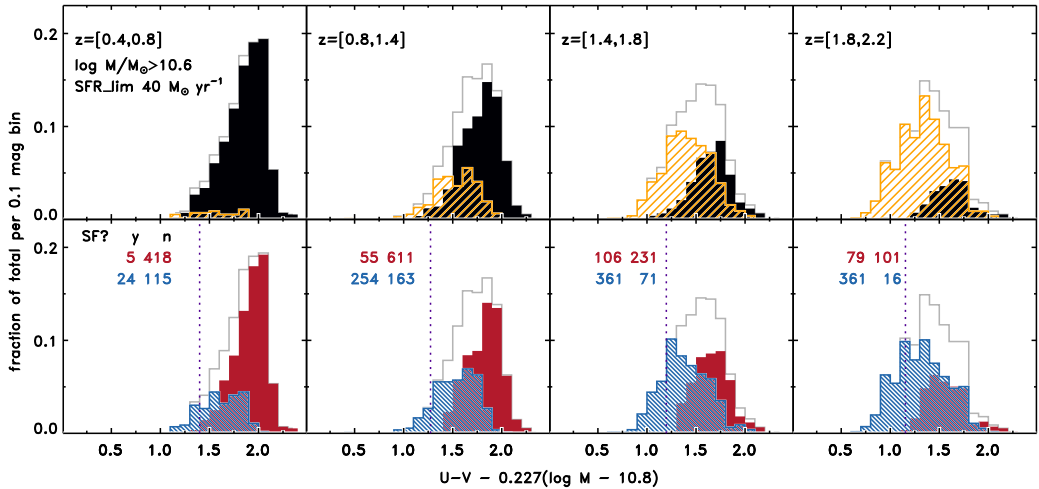}
	\caption[Rest-frame color histograms at $0.4 < z < 2.2$]{Rest-frame $U-V$ color distributions, with $U-V$ corrected for the (non-evolving) slope of the color-mass relation from \cite{borch:06} (see Fig. \ref{f:red_sequence}).  The histograms are normalized by the total number of galaxies in each redshift bin.  The distribution for all galaxies with $\log M/M_\odot > 10.6$---slightly lower than our completeness limit at $z=2$---is shown in the thin, grey histograms.  The top panels show the color distribution split between sources with $\SFRuvir$ greater (orange, hatched) or less (black, solid) than $40~M_\odot~\peryr$.  The bottom panels show the color distribution for red/blue galaxies selected as in Fig. \ref{f:color_color}.  The vertical dashed line in the bottom panels indicates the \cite{borch:06} red sequence selection limit, extrapolated to $z=2$.  The inset numbers show the number of galaxies in the ``quiescent'' and ``star-forming'' samples that have $\SFRuvir$ greater (``y'') or less (``n'') than $40~M_\odot~\peryr$.\label{f:histograms}}
\end{figure*}
%%%%%%%%%%%%%%%%%%%%%%%%%%%%%%%%%%%%%%%%%%%%%%%%%%%%%%%%%%%%%%%%%%%

We first analyze simple color-mass relations as a function of redshift.
As we show below, the use of a single color to identify different
galaxy populations is sufficient at low redshift but is not very
meaningful at redshifts $z>1$. The
$\UmV$ color-mass relations in four
redshift bins are shown in Fig.\ \ref{f:red_sequence}.
The redshift bins are chosen such that they sample similar comoving
volumes, with the exception of the lowest-redshift bin that
encompasses a volume only 40\% of the other bins.

Areas of the color-mass diagram where the
sample is incomplete are shaded on Fig. \ref{f:red_sequence}.
Selection in the $K$-band is closer to a stellar mass
selection than an optical selection is
\citep[e.g.,][]{franx:08}; nevertheless,  a fixed $K$-band flux
limit will result in a mass completeness limit that varies as a
function of redshift and color \citep{taylor:09a, marchesini:09}.
The completeness limits were estimated
from the NMBS sample itself combined with
the deeper FIREWORKS catalog of the CDFS \citep{wuyts:fireworks}
following the method described by \cite{taylor:09a} and
\cite{marchesini:09}.  At
$K<22.8$ we are complete for all galaxies with $M>10^{11}~M_\odot$ at
$0 < z < 2.2$.  The completeness limit extends to
$\sim10^{10}~M_\odot$ for blue galaxies with lower $M/L$ ratios.  We
avoid uncertain completeness corrections below by only considering
stellar mass ranges above the completeness threshold at a given
redshift.

The top panels of
Fig. \ref{f:red_sequence} highlight a key aspect of
massive galaxies at high redshift, which has earlier been
highlighted by
others \citep[e.g.,][]{papovich:06,brammer:09,whitaker:10}: nearly all
galaxies with $M > 10^{11}~M_\odot$ have red $\UmV$ colors at $0.4 < z
< 2.2$. However, this does not imply that all
these galaxies are quiescent with
low star formation rates. In the bottom panels of
Fig.\ \ref{f:red_sequence}
galaxies with $\SFRuvir>40\ M_\odot~\peryr$ are shown in
orange.  This star formation rate limit corresponds to the
minimum SFR observable at $z=2.2$ given the MIPS flux limit of
$20~\uJy$.  A significant fraction of red galaxies with $M>10^{11}M_\odot$
are vigorously forming stars at $z>1$, with the fraction increasing
with redshift and reaching $\sim$50\% by $z=2$ \citep{whitaker:10}.
This is qualitatively
consistent with the observed increase in the number of
IR-luminous galaxies with redshift \citep{lefloch:05, lefloch:09}.  In
the low-$z$ bin, galaxies on the massive red sequence have much lower
star formation rates, which is consistent with many other studies that
have found that red galaxies at $z<1$ tend to be old,
passively-evolving, early-type systems
\citep[e.g.][]{strateva:01,blanton:03,kauffmann:03,bell:04}.
%Red
%galaxies with SFRs less than the limit in Fig. \ref{f:red_sequence}
%follow a color-mass relation up to $z=2$: the average color becomes
%redder with increasing mass at a given redshift.  Furthermore, the
%color zeropoint of the red sequence becomes bluer with increasing
%redshift at a rate roughly consistent with an extrapolation of the
%$z<1$ color evolution found by \cite{bell:04} and \cite{borch:06} for
%the COMBO-17 survey.

%\subsection{Dust-corrected $\UmV$ colors}\label{s:dust_correction}
\subsection{Separating quiescent and star-forming galaxies}\label{s:separation}

Given that red galaxies at $z>1$ comprise both
quiescent galaxies and dusty star-forming galaxies,
a single rest-frame $U-V$ color criterion is not the most informative
way to identify different galaxy populations.
Recent studies of the local universe and of galaxies up to $z\sim1$
have shown that the galaxy color bimodality is more clearly seen after
accounting for dust reddening \citep{wyder:07, cowie:08, maller:09},
which \cite{brammer:09} extend to $z\sim2.2$ using the NMBS.
Quiescent galaxies follow a red sequence at least up to $z=2.2$
\citep{brammer:09, whitaker:10}, while star-forming galaxies follow a
sequence where the $\UmV$ colors become redder with increasing mass,
with the color primarily determined by increasing dust reddening with
mass \citep{labbe:07, brammer:09}.

\cite{labbe:05} show that it is possible to separate quiescent from
dusty galaxies with similarly red colors using the combination of
\textit{two} rest-frame colors, $\UmV$ and $\VmJ$ \citep[see
  also][]{wuyts:07, williams:09}.  The SEDS of quiescent galaxies are
red in $\UmV$ but blue beyond the Balmer/4000~\AA\ break, while
dust-reddened galaxies are red in both colors.  We choose here to use
this two-color ``$UVJ$'' selection technique because it is independent
of differences between population synthesis models, while the
dust-corrected $\UmV$ selection used by \cite{brammer:09} depends on
accurately measuring $A_V$ from the SED fit.  \cite{whitaker:10}
compare the $UVJ$ and dust-corrected color selection methods and find
that the fraction of red/quiescent galaxies with $M>10^{11}~M_\odot$
differs by less than 5\% between them.

%%%%%%%%%%%%%%%%%%%%%%%%    Mass functions   %%%%%%%%%%%%%%%%%%%%%    
\begin{figure*}
	\epsscale{1.1}
	\plotone{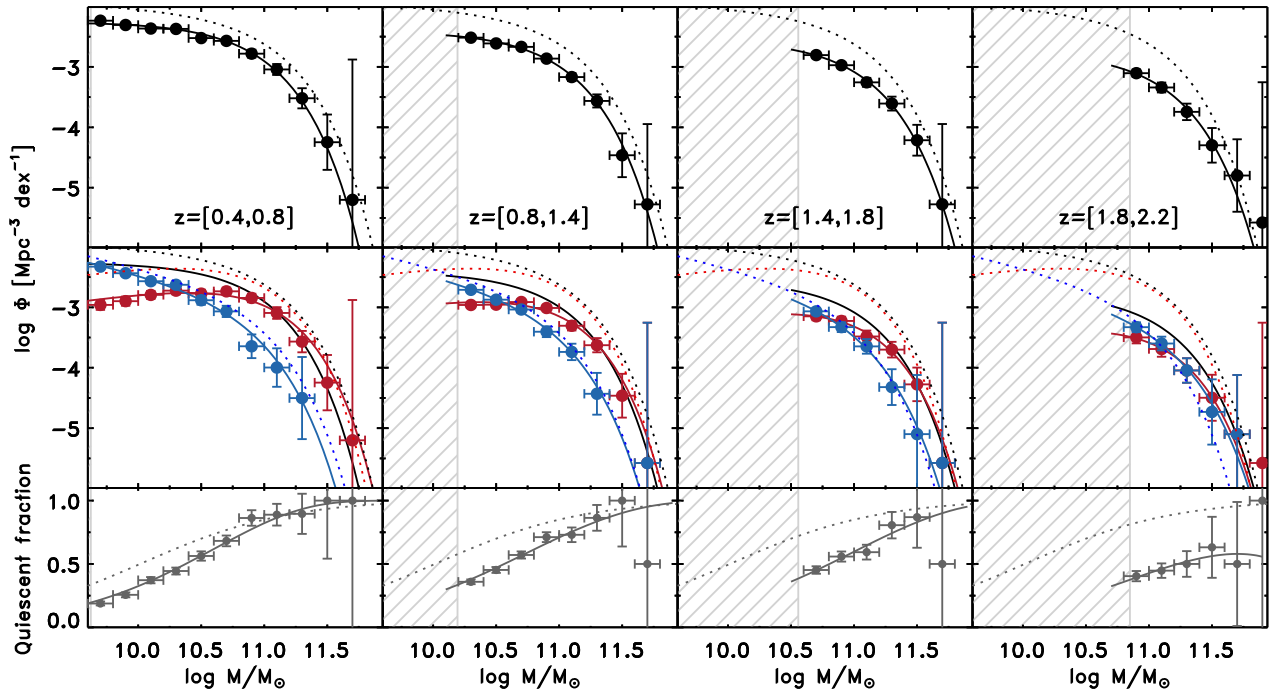}

	\caption[Stellar mass functions for the full sample and the
          quiescent/star-forming subsamples]{Stellar mass
functions for the full
          NMBS sample (top panels) and split using the quiescent/star-forming
          selection shown in Fig. \ref{f:color_color} (middle
          panels).  The points shown are simple redshift histograms
          divided by the volume of the NMBS, with Poisson error-bars.
          Representative \cite{schechter:76} function fits are shown,
          with the rest-frame slope fixed to $\alpha=$-0.99, -1.4, and
          -0.7 for the full, star-forming, and quiescent samples, respectively.  The
          dotted lines show the local stellar mass function of all
          (black), early- (red) and late-type (blue) galaxies
          \citep{bell:03}, scaled as described in the text. The light
          hatched regions show the 90\% completeness limit for red
          galaxies at the high-redshift end of each bin.  Note that we
          determine number and mass densities below by simply counting
          objects at masses where the NMBS is complete, rather than
          integrating the Schechter functions.  The bottom panels show
          the fraction of red, quiescent galaxies as a function of
          stellar mass and redshift.  The dotted line shows the ratio
          of the \cite{bell:03} early and early+late Schechter
          functions, while the solid lines show the ratio of the
          Schechter function fits to the NMBS mass
          functions. Quiescent galaxies clearly evolve much
more rapidly than star-forming galaxies, driving the evolution of the
total mass function at the high mass end. Here and in the figures below, all galaxies are shown whether or not they have MIPS coverage.  \label{f:mass_functions}}

\end{figure*}
%%%%%%%%%%%%%%%%%%%%%%%%%%%%%%%%%%%%%%%%%%%%%%%%%%%%%%%%%%%%%%%%%%%%

Figure \ref{f:color_color} shows the $\UmV$ and $\VmJ$ colors of
galaxies in the NMBS at $0.4 < z < 2.2$.  The two-color distribution
is bimodal up to $z=2.2$ \citep[see also][]{williams:09, ilbert:10},
and we adopt the \cite{williams:09} selection criteria (solid line) to
separate quiescent galaxies
and star-forming galaxies.  Galaxies with $\SFRuvir >
40~M_\odot\mathrm{yr}^{-1}$ are again shown with orange symbols in
Fig.\ \ref{f:color_color}.  While this SFR limit, set by the depth of
the \24mum observations, is insufficient for firmly
establishing a galaxy to be ``quiescent'', the clear separation in the
$UVJ$ diagram supports the idea  that the quiescent red galaxies form a
population distinct from the dusty, star-forming galaxies,
particularly at $z>1.5$.  We note here that
the existence of massive galaxies
with very little ongoing star-formation at $z\sim 2$ is supported
by near-IR spectroscopic studies \citep{kriek:06b}.

It is clear that red (quiescent) and blue (star-forming) samples selected using the $UVJ$ colors
will be different from those based on a single rest-frame color,
as was done by, e.g., \cite{bell:04} and \cite{borch:06}.  This is
shown qualitatively in Fig.\ \ref{f:histograms}, which shows histograms
of $\UmV$ color corrected for the slope of the color-magnitude
relation.  The top panels show the color distribution split according
to the SFR threshold used in Fig. \ref{f:red_sequence}, while the
bottom panels show the distribution split according to the $UVJ$
quiescent/star-forming selection. Although
the distribution of $U-V$ colors changes only slightly with redshift,
the distribution of star-forming versus quiescent galaxies changes
rapidly with redshift.
Hereafter, we
discuss the two distinct populations apparent in the $UVJ$ diagram as
``quiescent'' and ``star-forming'' samples. We cannot exclude some
on-going star formation in the quiescent galaxies (particularly
if it is completely obscured), but their rest-frame optical SEDs are
dominated by an evolved stellar population.

It is worth commenting here on other recent studies of the evolution
of the red sequence to $z\sim 2$.  A number of studies have found
little or no evidence for a significant number of red, quiescent
galaxies at $z>1.5$ \citep{arnouts:07, cirasuolo:07}.  Like the NMBS,
these studies rely on photometric redshifts, but they have relatively
poor sampling of observed NIR wavelengths with only the $J$ and $K$
broad-band filters.  This will result in large redshift uncertainties
at $z=1.5\mbox{--}2$ \citep{brammer:08}, and therefore derived
rest-frame colors will have insufficient precision to identify the
distinct red/blue populations at those redshifts \citep{taylor:09a}.
Though sample sizes are small, spectroscopic studies have unambiguously
identified quiescent galaxies at $z\sim 2$ \citep{cassata:08, kriek:08},
emphasizing the need for precise redshift measurements.  Furthermore,
the definition of ``quiescence'' varies from one study to another.
\cite{arnouts:07} define galaxies to be quiescent when they are
best-fit by a non-evolving (local) elliptical galaxy template. As
can be seen in
Fig.\ \ref{f:red_sequence},
passive evolution ensures that quiescent galaxies at $z>1.5$ cannot
have the colors of local elliptical galaxies, and thus such a
selection will fail to identify this population.

%%%%%%%%%%%%%%%%%%%%%%%%%%%%%%%%%%%%%%%%%%%%%%%%%%%%%%%%%%%%%%%%%%
%
%   Mass functions
%
%%%%%%%%%%%%%%%%%%%%%%%%%%%%%%%%%%%%%%%%%%%%%%%%%%%%%%%%%%%%%%%%%%
\section{Stellar mass functions}\label{s:mass_functions}

%%%%%%%%%%%%%%%%%%%%%%%%    Densities integrated to mass limit   %%%%%%%%%%%%%%%%%%%%%    
\begin{figure*}
	\plotone{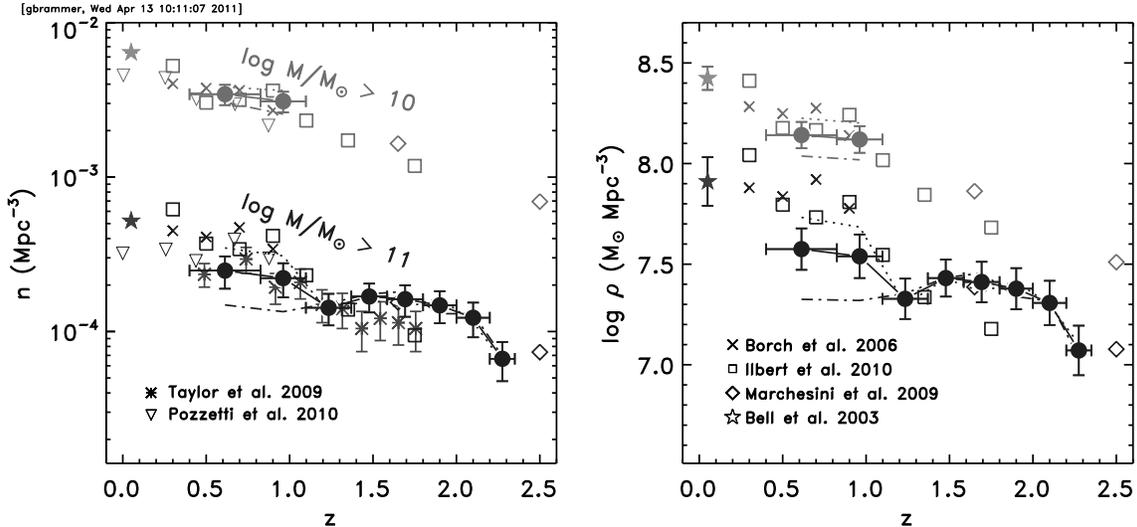}

	\caption{\textit{(Left panel)} Number density
          evolution of galaxies in the NMBS selected to mass limits of
          $\log M/M_\odot > 10, 11$ (large filled circles with error bars).  The redshift error-bars show the
          width of the redshift bin used, which are half the width of
          those used in
          Figs.\ref{f:red_sequence}--\ref{f:mass_functions}.  The
          density errors include Poisson and cosmic variance errors
          computed following \cite{somerville:04b}, added in
          quadrature.  The densities for the individual AEGIS and
          COSMOS fields are shown with dotted and dash-dotted lines,
          respectively.  \textit{(Right panel)} Mass density
          integrated to the mass limits indicated in the left panel.
          The errors shown are the fractional errors on the number
          densities, and do not include systematic uncertainties
          associated with the stellar-mass determinations.  Number and
          mass densities integrated from the mass functions of
          \cite{borch:06,marchesini:09,ilbert:10}, scaled to match our
          SED modeling assumptions where necessary, are shown with
          symbols as labeled.  Number densities are taken directly
          from \cite{taylor:09a} and
          \cite{pozzetti:10}.  The numerical values of the densities and total errors for the full NMBS sample are provided in Table \ref{t:tab1} below. \label{f:density_mass_limit}}

\end{figure*}
%%%%%%%%%%%%%%%%%%%%%%%%%%%%%%%%%%%%%%%%%%%%%%%%%%%%%%%%%%%%%%%%%%%%%%

Having robustly divided the sample
in quiescent and star-forming galaxies, we now study the evolution
of the stellar mass function split by galaxy type.
We first show the
evolution of all galaxies,
in the top panels of
Fig. \ref{f:mass_functions} \citep[see also, e.g.,][]{fontana:06, marchesini:09, pozzetti:10}. These
mass functions are computed by simply counting galaxies in stellar
mass and redshift bins.  We do not adopt the $V_\mathrm{max}$
formalism \citep{avni:80} as we only consider stellar masses where the
NMBS is complete.  We fit \cite{schechter:76} functions with fixed
faint-end slope to the densities in each redshift bin to demonstrate
only that the mass functions have reasonable shapes.  Our stellar mass
completeness limits do not allow us to constrain the faint-end slope
at $z>1$, and strong degeneracies between the Schechter parameters
\citep[e.g.][]{marchesini:09} would make the parameter values and
their evolution with redshift difficult to interpret given the simple
analysis used here.  We defer a more detailed analysis of the NMBS
stellar mass functions, including a full accounting of systematic
errors and incompleteness, to a future paper.

We include in Fig. \ref{f:mass_functions} the $z=0.1$ mass function
from \cite{bell:03} for comparison, which we have scaled to our
assumed cosmology.  Additionally, we scale the \cite{bell:03} stellar
masses down by a factor of 1.2 to account for the difference between a
``diet-Salpeter'' IMF and the \cite{kroupa:01} IMF we use to estimate
stellar masses.  It is apparent from the total mass functions that the
mass function evolves gradually from $z=0$ to $z=2$, with no indications
for sudden or dramatic changes in particular redshift ranges.

The mass functions are split into quiescent and star-forming galaxies
in the
middle panels of Fig. \ref{f:mass_functions}.
The dotted lines indicate the \cite{bell:03} local
mass functions of early- and late-type populations separated by color.
It is clear that the mass
functions of the quiescent and star-forming samples evolve in significantly
different ways.  Massive galaxies ($M>10^{11}M_\odot$) at $z<1$ are almost
entirely in the quiescent population.  
The number of massive quiescent
galaxies decreases steadily with increasing redshift, while the mass
function of the star-forming sample evolves very little up to $z=2$ for
$M\gtrsim10^{10.8}~M_\odot$.  The differential evolution is such that
the two populations are approximately equal in number at $z=2$ for these galaxies with $M>10^{10.8}M_\odot$ (Fig. \ref{f:mass_functions}, bottom panels).  Thus
we observe in the galaxy population directly the argument made by
\cite{bell:07}, who noted that the IR-luminous (massive) starburst
galaxies observed at $z\gtrsim2$ must later migrate to the red
sequence to avoid dramatically over-predicting the number of massive,
blue galaxies observed locally. Furthermore, we demonstrate here that
the evolution of the total galaxy population is mostly due to the evolution
of the quiescent galaxies: the star-forming population is similar
at all redshifts, and it is the rise of quiescent population
with cosmic time that is responsible for the gradual evolution
of the total mass function.

The shape of the quiescent mass function may also evolve with
redshift, although we are severely hampered by incompleteness
at low masses and high redshifts. Taking the data at face
value, we find that massive quiescent galaxies above $10^{11}M_\odot$ evolve more slowly than those at lower masses, but deeper data
at the low mass end and
spectroscopic redshifts at the high mass end are needed to confirm
this. We note that
this evolution would be qualitatively
consistent with the recent ($z<1$) build-up of low-mass red galaxies found
by \cite{rudnick:09}, both in cluster and field populations.

%%%%%%%%%%%%%%%%%%%%%%%%%%%%%%%%%%%%%%%%%%%%%%%%%%%%%%%%%%%%%%%%%%
%
%   Number and stellar-mass densities
%
%%%%%%%%%%%%%%%%%%%%%%%%%%%%%%%%%%%%%%%%%%%%%%%%%%%%%%%%%%%%%%%%%%
\section{Number and mass density evolution}\label{s:density}

\subsection{Full galaxy sample}

%%%%%%%%%%%%%%%   Number/mass density evolution in mass bins %%%%%%%%%%%%%%%%%%%%%   

\begin{figure*}
	\plotone{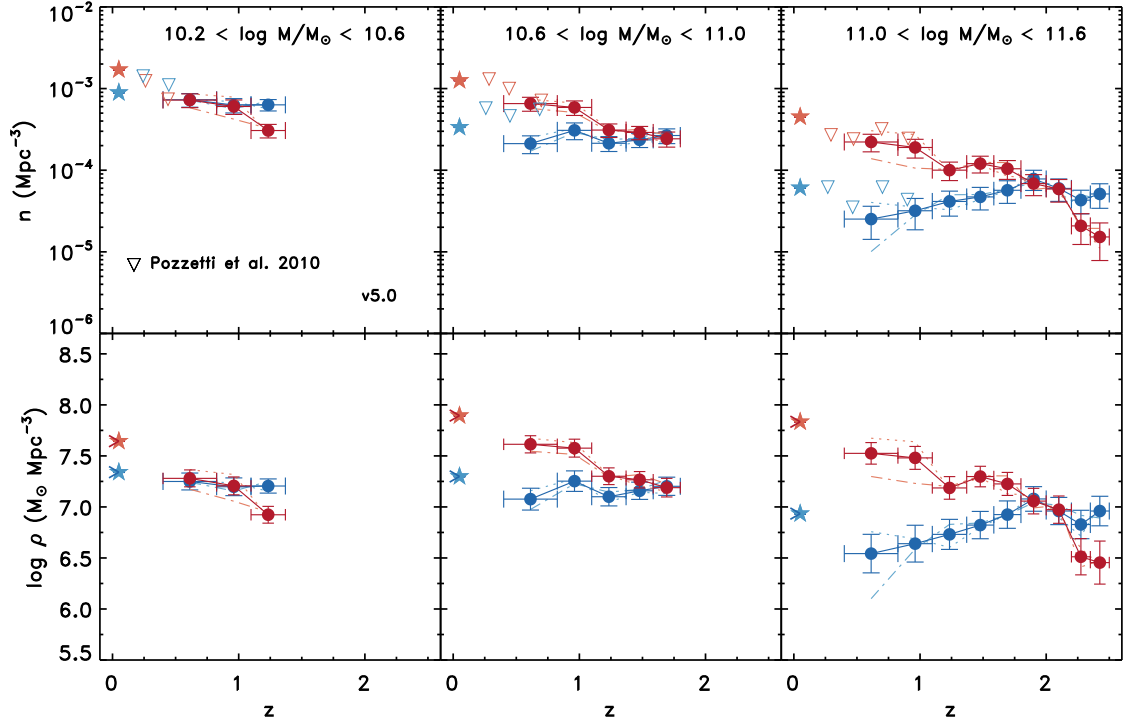}
	\caption{Number and mass density evolution for galaxies divided into quiescent and star-forming samples as in Fig. \ref{f:color_color} for three stellar mass bins.  Redshift bins are only shown where the mass bins are $>90\%$ complete.  The $z=0.1$ density measurements are integrated from the early- and late-type \cite{bell:03} mass functions (see Fig. \ref{f:density_mass_limit}.  Number densities of the ``red'' and ``blue'' types defined by \cite{pozzetti:10} are shown with open triangles, taken from their Fig. 13.  Note that the \cite{pozzetti:10} mass bins are slightly different than those used here ($10^{10-10.5}$, $10^{10.5-11}$, $10^{11-11.5}M_\odot$).  The numerical values of the densities and total errors for the full NMBS sample, divided by galaxy type, are provided in Table \ref{t:tab2} below. \label{f:dead_sequence_mass_bins}}

\end{figure*}

%%%%%%%%%%%%%%%%%%%%%%%%%%%%%%%%%%%%%%%%%%%%%%%%%%%%%%%%%%%%%%%%%%%%%%

To further quantify the evolution of star-forming and
quiescent galaxies, we now
consider the evolution of their number and mass densities with
redshift.  We measure these quantities directly from the data by
simply counting objects above the mass completeness threshold in a
given redshift bin, so we do not rely on the assumption that the mass
functions follow a Schechter function.  Again, the NMBS does not
constrain the faint-end of the mass function at $z>1$, so integrating
the mass functions to zero mass would result in an unreliable
extrapolation of the observed data.

The number and mass density evolution in two mass ranges is shown in
Fig.\ \ref{f:density_mass_limit}.  The redshift bins used are those
from Figs.\ \ref{f:red_sequence}--\ref{f:mass_functions} divided in
two; the bins are still somewhat larger than the expected photometric
redshift uncertainties even at $z=2$ (\S\ref{s:photoz}).  Only
redshift bins where the survey is complete to the specified mass
limits are shown.  The uncertainties on the number densities are a
combination of Poisson errors and the cosmic variance estimated using
the \cite{somerville:04b} prescription, given the observed number
density, redshift binning, and survey geometry.  We can obtain a rough
estimate of the cosmic variance directly from the NMBS itself: the
number and mass densities for the individual COSMOS and AEGIS fields
are shown as the dotted and dash-dotted lines, respectively.  The
differences between the two fields are significant and are generally
well-represented by the error estimates that account for cosmic
variance.

The inclusion of the local density measurement is important for
evaluating the overall evolution, and here we include densities
integrated from the \cite{bell:03} mass function.  While the
methodology we use to estimate stellar masses and separate quiescent
from star-forming
galaxies is quite different than that used by \cite{bell:03},
the local reference point appears to connect well with our NMBS
measurements in Fig. \ref{f:density_mass_limit}.  Similar to other studies, we find that there is modest mass density
evolution (0.25 dex) up to $z=1$ for masses $M>10^{10}M_\odot$.
Up to $z=2$, we observe an overall decrease in the mass density of
$\sim$0.6 dex at $M > 10^{11}~M_\odot$. We do not find a sudden
change at $z>1$; the evolution appears to be gradual over the
entire redshift range $0<z<2.2$.  We note that the results here are consistent with those of \cite{marchesini:09}, who find very little evolution at extreme masses $>10^{11.5}~M_\odot$ (Figure \ref{f:mass_functions}).

Differences between the two NMBS fields highlight
the importance of sampling large volumes:
the COSMOS field shows an overall mass
density decrease of 0.6 dex between $0.4 < z < 2$, while densities
measured in the AEGIS field are essentially consistent with
\textit{no} evolution over this same redshift range.  Furthermore,
we note that our results are sensitive to subtle redshift-dependent
systematic errors in the masses, including systematic differences between
the $z=0$ point and the higher redshift data. This is demonstrated
explicitly in Fig.\ \ref{f:mass_errors} in the Appendix.

\subsection{The number and mass densities of
quiescent and star-forming galaxies}\label{s:red_blue_density}

We examine the number and mass density evolution of the quiescent
and star-forming
samples (\S\ref{s:separation}) in
Fig. \ref{f:dead_sequence_mass_bins}.  Rather than integrating down to
a mass limit as in Fig. \ref{f:density_mass_limit}, we now consider
three separate stellar mass bins where the NMBS is complete to at
least $z=1$.  While this is essentially the same measurement as the
stellar mass functions described above, plotting the densities allows
the trends with redshift and mass to be more readily apparent.

It is immediately clear again that the quiescent and star-forming galaxy populations
evolve in very different ways.  The number and mass densities of
star-forming galaxies remain nearly constant with redshift for all
masses $\log M/M_\odot > 10.2$.  Star-forming
galaxies with $\log M/M_\odot >
11$, which are relatively rare at low redshift, are more prominent at
$z=2$ where they have the same number and mass densities as the
massive quiescent population.  We again take the local comparison from
\cite{bell:03}, who provide stellar mass functions divided into early-
and late-type populations by color.  The local comparison is important
for showing the overall redshift evolution, particularly at
low masses where our data
indicate
rapid growth of the quiescent
population since $z\sim1$.  Whether or not the
local comparison is considered, the density evolution of star-forming
galaxies is in stark contrast with that of quiescent galaxies, whose
density
decrease with redshift at all of the masses considered.  The
evolution of the mass density of the quiescent galaxies as a function of redshift is
to first order
independent of mass for $\log M/M_\odot > 10.2$, with about
0.5 dex of mass density growth per unit redshift for all of the mass
ranges shown.  Because three times as much time elapses between $z=1$ and the present day as between $z=2$ and $z=1$, the massive red sequence grows rapidly
between $z=2$ and $z=1$, at which point the growth slows considerably
over the remaining $\sim$8 Gyr until the present day \citep[see
  also][]{kriek:08}.

The differential density evolution between the quiescent and star-forming
populations shown in Fig. \ref{f:dead_sequence_mass_bins} provides
another manifestation of the ``downsizing'' phenomenon, in which the
characteristic sites of star-formation shift to lower mass galaxies at
later times (\citealp{cowie:96}, followed by e.g., \citealp{bell:05,
  juneau:05, bundy:06} and many others).  Here we see that the number
and mass densities of quiescent and star-forming galaxies are equal at
earlier times for increasing galaxy stellar mass, after which the
densities are dominated by the quiescent population.  This is
effectively a reformulation of taking the stellar mass at which the
mass functions cross to be the characteristic mass of transforming
galaxies from the star-forming to quiescent population \citep{bundy:06}.  Indeed, we
see a similar trend in the NMBS stellar mass functions in
Fig. \ref{f:mass_functions}: the quiescent and
star-forming stellar mass functions
cross at stellar masses that increase with redshift.

%%%%%%%%%%%%%%%%%%%%%%%%%%%%%%%%%%%%%%%%%%%%%%%%%%%%%%%%%%%%%%%%%%
%
%   Discussion
%
%%%%%%%%%%%%%%%%%%%%%%%%%%%%%%%%%%%%%%%%%%%%%%%%%%%%%%%%%%%%%%%%%%
\section{Discussion: The build-up of quiescent galaxies}\label{s:discussion}

%%%%%%%%%%%%%%%   Mass at fixed n %%%%%%%%%%%%%%%%%%%%%
\begin{figure}
	\epsscale{1.15}
	\plotone{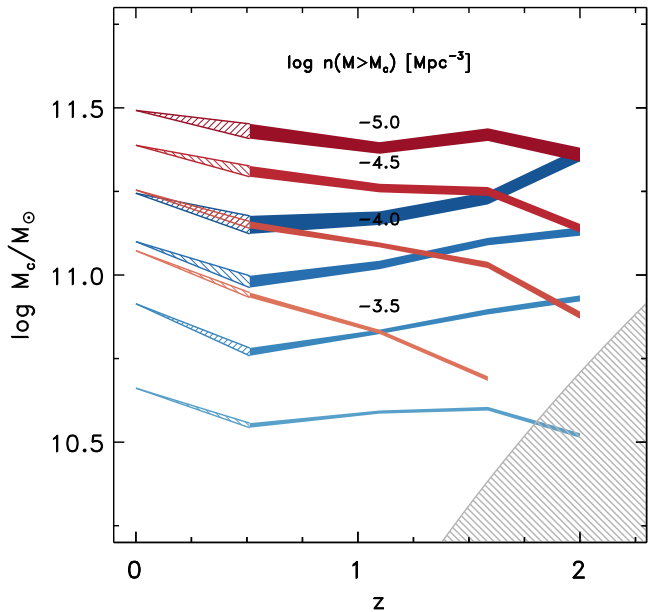}
	\caption[Redshift evolution of stellar mass at fixed cumulative number density]{Redshift evolution of stellar mass at fixed cumulative number density, $n(M>M_c)$, for the quiescent and star-forming galaxy samples. The uncertainty on $M_c$, estimated from Monte-Carlo simulations in which we perturb the stellar masses with a random error of 0.1 dex and remeasure the derived quantity, is indicated by the width of each color/density track.  The tracks are extended to $z=0$ using the \cite{bell:03} mass functions.  The shaded region at z=0 demonstrates the extent of the systematic uncertainty related to the model library used to derive stellar masses.  The gray shaded region shows the completeness limit of the NMBS.	\label{f:mass_at_fixed_n}}
\end{figure}
%%%%%%%%%%%%%%%%%%%%%%%%%%%%%%%%%%%%%%%%%%%%%%%%%%%%%%%%%%%%%%%%%%%%%%

The trends for star-forming and quiescent galaxies are contrary
to naive expectations.
One would naively
expect that the stellar mass density of star-forming galaxies
increases with time, as they are forming new stars.
Quiescent galaxies have already stopped forming
stars, and one might expect their mass density to remain constant.
However, we observe the opposite.
Consistent with previous studies
at lower redshift \citep{borch:06, bell:07, martin:07}, we
find that the mass density of the star-forming population shows
very little evolution up to $z\sim 2$ for galaxies with
$\log M/M_\odot\sim10.6$. At the same time, the mass density in
quiescent galaxies increases with time. The obvious interpretation
is that galaxies migrate from the star-forming population to
the quiescent population
\citep[see also, e.g.,][]{bell:07, faber:07}. Furthermore, the
mass density of the massive quiescent population above $10^{11}M\odot$ might grow through
mergers with less massive galaxies
\citep[e.g.,][]{vandokkum:05}.

\subsection{Mass evolution at fixed cumulative
number density}
\label{s:buildup}

To explore the causes of the rise of quiescent galaxies
we first ask how much the masses of individual galaxies grow with time.
As shown by \cite{vandokkum:10}, selecting
galaxies at a constant number density rather than a constant mass
enables the study of the evolution with time of a single coherent population of galaxies at the massive, exponential end of the mass function, as both
star formation and any merger with a ratio somewhat less than than 1:1 will
increase these galaxies' masses without changing their number density. 
\cite{vandokkum:10} find that 
galaxies with a number density
$n=2\times{10}^{-4}~\mathrm{Mpc}^{-3}~\mathrm{dex}^{-1}$ grow in mass
by a factor of two since $z=2$ (with $\log M/M_\odot = 11.15$ at
$z=2$).

%%%%%%%%%%%%%%%%     Figure: massive galaxy density evolution, linear units %%%%%%%%%%
\begin{figure*}
	\epsscale{1.15}
	\plotone{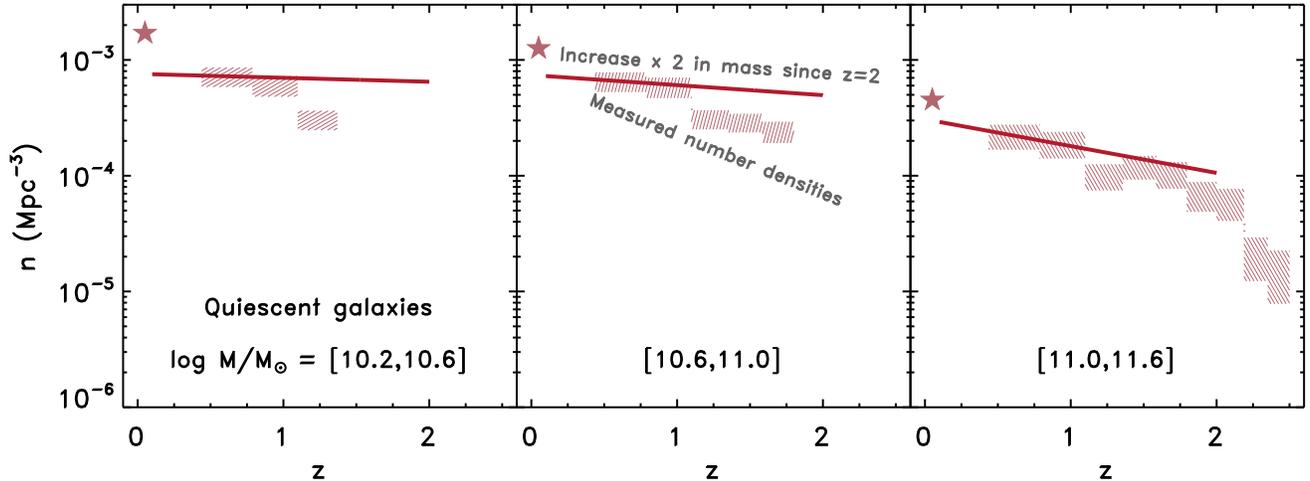}
	\caption{Number density evolution of quiescent galaxies.  The shaded regions are the same number densities and their uncertainties for quiescent galaxies as in Fig.  \ref{f:dead_sequence_mass_bins}.  The stars indicate the $z=0.1$ measurement from \cite{bell:03}.  The solid lines indicate the number density evolution produced by a very simple calculation in which we assume that all quiescent galaxies grow in mass by a factor of two from $z=2$ to $z=0$ (\S\ref{s:buildup}). \label{f:mf_deriv}}
\end{figure*}
%%%%%%%%%%%%%%%%%%%%%%%%%%%%%%%%%%%%%%%%%%%%%%%%%%%%%%%%%%%%%%%%%%%%%%

We modify the \cite{vandokkum:10} selection slightly and consider a
selection based on a constant \textit{cumulative} number density,
defining $M_c$ to be the stellar mass at constant $n(M>M_c)$.  This
definition has the advantage that it is single-valued even at
low-masses.  Starting with a baseline mass function with $M_c =
M_{c,0}$ at time $t=0$, the value of $M_c$ can increase after time
$t$ (i.e., $M_{c,t} > M_{c,0}$) as a result of three physical process
that all may occur simultaneously: growth from mergers where the
secondary objects have $M<M_c$ or both progenitors have $M\approx M_c$, star formation, and creation of
new galaxies in the considered sample via transformation of galaxies
(with $M>M_c$) from a different population.  In the first two of these
processes, an increase in $M_{c,t}$ is directly related to the average
increase in mass in individual galaxies above that threshold, given
the simplifying assumption that all galaxies above the threshold
evolve in a uniform way.  If two galaxies each with $M > M_{c,0}$
merge within time, $t$, then $M_{c,t} < M_{c,0}$, as one would have to
move further down the mass function to count the same number of
galaxies in the same volume.  Such mergers with both progenitors with $M > M_{c,0}$ would necessarily have mass ratios of nearly 1:1
for $M_c\gtrsim10^{11}~M_\odot$ due to the steepness of the
mass function.  Finally, mass loss in galaxies above $M_{c,0}$ after some time, $t$, will result in $M_{c,t} < M_{c,0}$.

We show the evolution of $M_c$ for the quiescent and star-forming
galaxy populations
in Fig. \ref{f:mass_at_fixed_n}, for a range of $n(M>M_c)$.  We find
that the stellar mass of quiescent galaxies with $\log
M/M_\odot\gtrsim11$ at $z=2$ increases by a factor of $\sim$2 by
$z=0$, similar to the \cite{vandokkum:10} value for the full sample
not divided by color.  Interestingly, the growth of $M_c$ at the
highest masses slows with time, perhaps suggesting an increased
prevalence of dry, equal-mass mergers at these masses since $z=1$.
The value of $M_c$ for the star-forming galaxies is nearly constant with
redshift for all densities considered, which is likely the result of
tension between competing and opposite effects: star-formation within
the population and migration out of the star-forming and into the
quiescent population.

\subsection{Mergers versus truncation of star formation}

We now construct a very simple empirical calculation to test the
hypothesis that the observed number density evolution of quiescent
galaxies is a result of growing the stellar mass of individual
galaxies by the factor of two measured from
Fig.\ \ref{f:mass_at_fixed_n}.  For the quiescent population, this
growth should be dominated by minor mergers since 1) these galaxies
are not forming stars at a high rate and 2) equal mass mergers are
rare, at least at the massive end, because the galaxies themselves are
relatively rare \citep[see also][]{vandokkum:10}.
The shape of the mass function ensures that there is
a large number of low-mass merger progenitors.  Taking the
\cite{marchesini:09} $1.3 < z < 2$ function, we measure the change in
number density in each mass bin that results from increasing all
masses by a factor of two.  This is essentially taking the derivative
of the mass function at the indicated mass bins.  Thus, the number
density evolution for a given increase in stellar mass is greatest in
the mass bin where the mass function is steepest.

The density evolution that is implied by this calculation is shown in Fig.
\ref{f:mf_deriv}, where we normalized the evolution the density observed in the
NMBS at $z=1$. The modest assumed growth of stellar mass in individual galaxies
results in a factor of 3 growth in the number density since $z=2$ for the
massive bin. This simple model can explain at least $\sim$50\% of the observed
density growth in the massive bin, suggesting that the growth in the number and
mass densities of these galaxies is dominated by (minor) mergers \citep[see
also][]{hopkins:10, vandokkum:10}. The remaining density growth of massive
quiescent galaxies can likely be explained by transforming galaxies from the
dusty star-forming population, which is indeed required to avoid overproducing
massive star-forming galaxies, as in Fig. \ref{f:mass_functions}

By contrast, growing the stellar masses of individual quiescent galaxies by a
factor of two can only account for 10--15\% of their observed number density
evolution at lower masses. If this modest mass growth within the less massive
quiescent population (e.g. due to red mergers) is insufficient to explain the
rapid increase of their number and mass densities, transformation of galaxies
from the star-forming to quiescent populations is likely the dominant mechanism
necessary to make up the difference. This conclusion is further supported by the
lack of evolution of the number/mass densities and $M_c$ for the star-forming
population as seen in Figs. \ref{f:dead_sequence_mass_bins} and
\ref{f:mass_at_fixed_n}, respectively.

\subsection{Consequences of the massive, dusty galaxy population at $z>1$}

The process by which galaxies transform from the star-forming sequence to the
quiescent sequence is unknown, but at moderate and high stellar masses it can
not be a simple fading of blue, dust-free star forming galaxies to red,
dust-free quiescent galaxies. As noted earlier, the bimodal galaxy population at
$z>1$ is not immediately apparent in the $\UmV$ color-mass diagram because the
star-forming and quiescent galaxies have essentially identical red colors (Fig.
\ref{f:red_sequence}). The former, we have argued, are heavily reddened by dust.
At no point for $z\lesssim2$ do we see a significant population of massive,
star-forming galaxies with un-reddened blue colors. This lack of observed blue
galaxies indicates that the dusty nature of the massive star-forming galaxies we
observe at $z>1$ is not likely to be a simple effect of the dust geometry---for
example that highly-inclined (blue) disks would have reddened colors that would
put them on the red sequence \citep[e.g.][]{bailin:08, maller:09}---because the
random orientation of objects would show a range of obscuration at constant
stellar mass broader than that observed.

As a result, massive galaxies must leave the star-forming sequence and
eventually arrive on the quiescent sequence at essentially the same $U-V$ color.
Although local early-type galaxies are frequently observed to contain modest
amounts of dust \citep{kormendy:89}, the dust does not appear to affect the
overall optical colors of nearby red-sequence galaxies to the dramatic extent
that we find in the optical-NIR SEDs of massive star-forming galaxies at $z>1$
\citep[see also][]{whitaker:10}. Therefore, dust must be destroyed or the dust
geometry needs to change during the process of of transforming galaxies from the
dusty star-forming sequence to the quiescent red sequence. As these processes
probably operate on timescales similar to or longer than the timescale of star
formation quenching, one might expect an increase in the amount of dust in the
quiescent population with redshift. \cite{marchesini:10} do find that extremely
massive galaxies at $z>3$ in the NMBS have fairly strong Balmer/4000~\AA\ breaks
and also $>1$ mag of dust extinction, measured with the same techniques
described in \S\ref{s:fast}. Future work combining medium-band surveys such as
the NMBS with new far-IR capabilities to better characterize emission from the
dust itself \citep[e.g.][]{elbaz:10}, will help to further test this prediction
of moderate dust in quiescent galaxies at $z>1$.

%%%%%%%%%%%%%%%%%%%%%%%%%%%%%%%%%%%%%%%%%%%%%%%%%%%%%%%%%%%%%%%%%%
%
%   Summary and conclusions
%
%%%%%%%%%%%%%%%%%%%%%%%%%%%%%%%%%%%%%%%%%%%%%%%%%%%%%%%%%%%%%%%%%%
\section{Summary}\label{s:summary}

We use the unique photometric dataset provided by the NEWFIRM Medium-Band Survey
to study the color and stellar mass distributions of galaxies up to $z=2$. The
NMBS provides the best-sampled SEDs and most precise photometric redshift
estimates at $z\gtrsim1$ available to date, which greatly improves the quality
of rest-frame properties (colors, masses, stellar population parameters) derived
from the photometry. We find that nearly all galaxies with $\log M/M_\odot >
10.5$ have red rest-frame $\UmV$ colors up to $z=2$. Using a two-color method,
supported by MIPS 24\micron\, photometry, we identify a bimodal galaxy
population consisting of a quiescent sequence with relatively low star-formation
rates and a distinct star-forming sequence that becomes increasingly dusty with
increasing stellar mass.

Separating the evolution of the mass function by star formation rate, we find
that it is driven by the rise of quiescent galaxies from $z=2$ to $z=0$. The
mass function of star-forming galaxies is remarkably similar at all redshifts,
whereas the quiescent galaxies show strong, mass-dependent, evolution \citep[see
also][]{marchesini:09, drory:09}. Quantifying this evolution, we find that
number and stellar mass densities of all galaxies with $\log M/M_\odot > 10$
evolve by $\sim$0.5 dex per unit redshift. Considering the density evolution of
star-forming and quiescent galaxies separately, we find that the density of
star-forming galaxies is nearly flat out to $z=2$, and the density of quiescent
galaxies decreases by a factor of $\sim 10$ from $z=0$ to $z=2$. Interestingly,
at $z=2$ the mass and number densities of the quiescent and star-forming
populations with $M>10^{11}~M_\odot$ are nearly identical. This crossing point
of the densities occurs at progressively later times at lower stellar masses.

Using an empirical argument based on selecting subsamples based on their
cumulative number density, we show that the average mass in individual quiescent
galaxies grows by a factor of $\sim$2 from $z=2$ to $z=0$. If we assume that
most of this mass growth is due to mergers, we find that a simple calculation
based on the shape of the mass function is able to explain much (at least
$\sim$50\%) of the number density evolution of galaxies with $M >
10^{11}M_\odot$. This model is unable to account for the density evolution of
less massive quiescent galaxies, which we argue are formed primarily via
transformations from the star-forming population. In general, our results bring
together a large variety of results from the literature---based on disparate
samples selected at different wavelengths and redshifts---into a coherent
picture extended up to $z=2$, thus demonstrating the utility of large,
uniformly-selected surveys for the study of galaxy formation and evolution.

%\clearpage

%%%%%%%%%%%%%%%%%%%%%%%%%%%%%%%%%%%%%%%%%%%%%%%%%%%%%%%%%%%%%%%%%%
%
%   End things
%
%%%%%%%%%%%%%%%%%%%%%%%%%%%%%%%%%%%%%%%%%%%%%%%%%%%%%%%%%%%%%%%%%%
\acknowledgements

Support from NSF grants AST-0449678 and AST-0807974 is gratefully acknowledged.  This research has made extensive use of the IDL Astronomy Library (\weblink{http://idlastro.gsfc.nasa.gov/}) and NASA's Astrophysics Data System Bibliographic Services.

{\it Facilities:} \facility{Mayall (NEWFIRM)}

\bibliographystyle{apj}
\bibliography{ms}

\begin{thebibliography}{91}
\expandafter\ifx\csname natexlab\endcsname\relax\def\natexlab#1{#1}\fi

\bibitem[{{Arnouts} {et~al.}(2007){Arnouts}, {Walcher}, {Le F{\`e}vre},
  {Zamorani}, {Ilbert}, {Le Brun}, {Pozzetti}, {Bardelli}, {Tresse}, {Zucca},
  {Charlot}, {Lamareille}, {McCracken}, {Bolzonella}, {Iovino}, {Lonsdale},
  {Polletta}, {Surace}, {Bottini}, {Garilli}, {Maccagni}, {Picat},
  {Scaramella}, {Scodeggio}, {Vettolani}, {Zanichelli}, {Adami}, {Cappi},
  {Ciliegi}, {Contini}, {de la Torre}, {Foucaud}, {Franzetti}, {Gavignaud},
  {Guzzo}, {Marano}, {Marinoni}, {Mazure}, {Meneux}, {Merighi}, {Paltani},
  {Pell{\`o}}, {Pollo}, {Radovich}, {Temporin}, \& {Vergani}}]{arnouts:07}
{Arnouts}, S., {Walcher}, C.~J., {Le F{\`e}vre}, O., {et~al.} 2007, \aap, 476,
  137

\bibitem[{{Avni} \& {Bahcall}(1980)}]{avni:80}
{Avni}, Y., \& {Bahcall}, J.~N. 1980, \apj, 235, 694

\bibitem[{{Bailin} \& {Harris}(2008)}]{bailin:08}
{Bailin}, J., \& {Harris}, W.~E. 2008, \apj, 681, 225

\bibitem[{{Baldry} {et~al.}(2004){Baldry}, {Glazebrook}, {Brinkmann},
  {Ivezi{\'c}}, {Lupton}, {Nichol}, \& {Szalay}}]{baldry:04}
{Baldry}, I.~K., {Glazebrook}, K., {Brinkmann}, J., {et~al.} 2004, \apj, 600,
  681

\bibitem[{{Barmby} {et~al.}(2008){Barmby}, {Huang}, {Ashby}, {Eisenhardt},
  {Fazio}, {Willner}, \& {Wright}}]{barmby:08}
{Barmby}, P., {Huang}, J.-S., {Ashby}, M.~L.~N., {et~al.} 2008, \apjs, 177, 431

\bibitem[{{Bell} {et~al.}(2003){Bell}, {McIntosh}, {Katz}, \&
  {Weinberg}}]{bell:03}
{Bell}, E.~F., {McIntosh}, D.~H., {Katz}, N., {et~al.} 2003, \apjs, 149, 289

\bibitem[{{Bell} {et~al.}(2005){Bell}, {Papovich}, {Wolf}, {Le Floc'h},
  {Caldwell}, {Barden}, {Egami}, {McIntosh}, {Meisenheimer},
  {P{\'e}rez-Gonz{\'a}lez}, {Rieke}, {Rieke}, {Rigby}, \& {Rix}}]{bell:05}
{Bell}, E.~F., {Papovich}, C., {Wolf}, C., {et~al.} 2005, \apj, 625, 23

\bibitem[{{Bell} {et~al.}(2004){Bell}, {Wolf}, {Meisenheimer}, {Rix}, {Borch},
  {Dye}, {Kleinheinrich}, {Wisotzki}, \& {McIntosh}}]{bell:04}
{Bell}, E.~F., {Wolf}, C., {Meisenheimer}, K., {et~al.} 2004, \apj, 608, 752

\bibitem[{{Bell} {et~al.}(2007){Bell}, {Zheng}, {Papovich}, {Borch}, {Wolf}, \&
  {Meisenheimer}}]{bell:07}
{Bell}, E.~F., {Zheng}, X.~Z., {Papovich}, C., {et~al.} 2007, \apj, 663, 834

\bibitem[{{Blanton} {et~al.}(2003){Blanton}, {Hogg}, {Bahcall}, {Baldry},
  {Brinkmann}, {Csabai}, {Eisenstein}, {Fukugita}, {Gunn}, {Ivezi{\'c}},
  {Lamb}, {Lupton}, {Loveday}, {Munn}, {Nichol}, {Okamura}, {Schlegel},
  {Shimasaku}, {Strauss}, {Vogeley}, \& {Weinberg}}]{blanton:03}
{Blanton}, M.~R., {Hogg}, D.~W., {Bahcall}, N.~A., {et~al.} 2003, \apj, 594,
  186

\bibitem[{{Borch} {et~al.}(2006){Borch}, {Meisenheimer}, {Bell}, {Rix}, {Wolf},
  {Dye}, {Kleinheinrich}, {Kovacs}, \& {Wisotzki}}]{borch:06}
{Borch}, A., {Meisenheimer}, K., {Bell}, E.~F., {et~al.} 2006, \aap, 453, 869

\bibitem[{{Brammer} {et~al.}(2008){Brammer}, {van Dokkum}, \&
  {Coppi}}]{brammer:08}
{Brammer}, G.~B., {van Dokkum}, P.~G., \& {Coppi}, P. 2008, \apj, 686, 1503

\bibitem[{{Brammer} {et~al.}(2009){Brammer}, {Whitaker}, {van Dokkum},
  {Marchesini}, {Labb{\'e}}, {Franx}, {Kriek}, {Quadri}, {Illingworth}, {Lee},
  {Muzzin}, \& {Rudnick}}]{brammer:09}
{Brammer}, G.~B., {Whitaker}, K.~E., {van Dokkum}, P.~G., {et~al.} 2009, \apjl,
  706, L173

\bibitem[{{Brown} {et~al.}(2007){Brown}, {Dey}, {Jannuzi}, {Brand}, {Benson},
  {Brodwin}, {Croton}, \& {Eisenhardt}}]{brown:07}
{Brown}, M.~J.~I., {Dey}, A., {Jannuzi}, B.~T., {et~al.} 2007, \apj, 654, 858

\bibitem[{{Bruzual} \& {Charlot}(2003)}]{bc:03}
{Bruzual}, G., \& {Charlot}, S. 2003, \mnras, 344, 1000

\bibitem[{{Bundy} {et~al.}(2006){Bundy}, {Ellis}, {Conselice}, {Taylor},
  {Cooper}, {Willmer}, {Weiner}, {Coil}, {Noeske}, \& {Eisenhardt}}]{bundy:06}
{Bundy}, K., {Ellis}, R.~S., {Conselice}, C.~J., {et~al.} 2006, \apj, 651, 120

\bibitem[{{Calzetti} {et~al.}(2000){Calzetti}, {Armus}, {Bohlin}, {Kinney},
  {Koornneef}, \& {Storchi-Bergmann}}]{calzetti:00}
{Calzetti}, D., {Armus}, L., {Bohlin}, R.~C., {et~al.} 2000, \apj, 533, 682

\bibitem[{{Capak} {et~al.}(2007){Capak}, {Aussel}, {Ajiki}, {McCracken},
  {Mobasher}, {Scoville}, {Shopbell}, {Taniguchi}, {Thompson}, {Tribiano},
  {Sasaki}, {Blain}, {Brusa}, {Carilli}, {Comastri}, {Carollo}, {Cassata},
  {Colbert}, {Ellis}, {Elvis}, {Giavalisco}, {Green}, {Guzzo}, {Hasinger},
  {Ilbert}, {Impey}, {Jahnke}, {Kartaltepe}, {Kneib}, {Koda}, {Koekemoer},
  {Komiyama}, {Leauthaud}, {Lefevre}, {Lilly}, {Liu}, {Massey}, {Miyazaki},
  {Murayama}, {Nagao}, {Peacock}, {Pickles}, {Porciani}, {Renzini}, {Rhodes},
  {Rich}, {Salvato}, {Sanders}, {Scarlata}, {Schiminovich}, {Schinnerer},
  {Scodeggio}, {Sheth}, {Shioya}, {Tasca}, {Taylor}, {Yan}, \&
  {Zamorani}}]{capak:07}
{Capak}, P., {Aussel}, H., {Ajiki}, M., {et~al.} 2007, \apjs, 172, 99

\bibitem[{{Cardamone} {et~al.}(2010){Cardamone}, {van Dokkum}, {Urry},
  {Taniguchi}, {Gawiser}, {Brammer}, {Taylor}, {Damen}, {Treister}, {Cobb},
  {Bond}, {Schawinski}, {Lira}, {Murayama}, {Saito}, \&
  {Sumikawa}}]{cardamone:10}
{Cardamone}, C.~N., {van Dokkum}, P.~G., {Urry}, C.~M., {et~al.} 2010, \apjs,
  189, 270

\bibitem[{{Cassata} {et~al.}(2008){Cassata}, {Cimatti}, {Kurk}, {Rodighiero},
  {Pozzetti}, {Bolzonella}, {Daddi}, {Mignoli}, {Berta}, {Dickinson},
  {Franceschini}, {Halliday}, {Renzini}, {Rosati}, \& {Zamorani}}]{cassata:08}
{Cassata}, P., {Cimatti}, A., {Kurk}, J., {et~al.} 2008, \aap, 483, L39

\bibitem[{{Chary} \& {Elbaz}(2001)}]{chary:01}
{Chary}, R., \& {Elbaz}, D. 2001, \apj, 556, 562

\bibitem[{{Cirasuolo} {et~al.}(2007){Cirasuolo}, {McLure}, {Dunlop}, {Almaini},
  {Foucaud}, {Smail}, {Sekiguchi}, {Simpson}, {Eales}, {Dye}, {Watson}, {Page},
  \& {Hirst}}]{cirasuolo:07}
{Cirasuolo}, M., {McLure}, R.~J., {Dunlop}, J.~S., {et~al.} 2007, \mnras, 380,
  585

\bibitem[{{Conroy} {et~al.}(2009){Conroy}, {Gunn}, \& {White}}]{conroy:09}
{Conroy}, C., {Gunn}, J.~E., \& {White}, M. 2009, \apj, 699, 486

\bibitem[{{Cowie} \& {Barger}(2008)}]{cowie:08}
{Cowie}, L.~L., \& {Barger}, A.~J. 2008, \apj, 686, 72

\bibitem[{{Cowie} {et~al.}(1996){Cowie}, {Songaila}, {Hu}, \&
  {Cohen}}]{cowie:96}
{Cowie}, L.~L., {Songaila}, A., {Hu}, E.~M., {et~al.} 1996, \aj, 112, 839

\bibitem[{{Dale} \& {Helou}(2002)}]{dale:02}
{Dale}, D.~A., \& {Helou}, G. 2002, \apj, 576, 159

\bibitem[{{Davis} {et~al.}(2003){Davis}, {Faber}, {Newman}, {Phillips},
  {Ellis}, {Steidel}, {Conselice}, {Coil}, {Finkbeiner}, {Koo}, {Guhathakurta},
  {Weiner}, {Schiavon}, {Willmer}, {Kaiser}, {Luppino}, {Wirth}, {Connolly},
  {Eisenhardt}, {Cooper}, \& {Gerke}}]{davis:deep2}
{Davis}, M., {Faber}, S.~M., {Newman}, J., {et~al.} 2003, in Presented at the
  Society of Photo-Optical Instrumentation Engineers (SPIE) Conference, Vol.
  4834, Society of Photo-Optical Instrumentation Engineers (SPIE) Conference
  Series, ed. P.~{Guhathakurta}, 161--172

\bibitem[{{Drory} {et~al.}(2009){Drory}, {Bundy}, {Leauthaud}, {Scoville},
  {Capak}, {Ilbert}, {Kartaltepe}, {Kneib}, {McCracken}, {Salvato}, {Sanders},
  {Thompson}, \& {Willott}}]{drory:09}
{Drory}, N., {Bundy}, K., {Leauthaud}, A., {et~al.} 2009, \apj, 707, 1595

\bibitem[{{Elbaz} {et~al.}(2010){Elbaz}, {Hwang}, {Magnelli}, {Daddi},
  {Aussel}, {Altieri}, {Amblard}, {Andreani}, {Arumugam}, {Auld}, {Babbedge},
  {Berta}, {Blain}, {Bock}, {Bongiovanni}, {Boselli}, {Buat}, {Burgarella},
  {Castro-Rodriguez}, {Cava}, {Cepa}, {Chanial}, {Chary}, {Cimatti},
  {Clements}, {Conley}, {Conversi}, {Cooray}, {Dickinson}, {Dominguez},
  {Dowell}, {Dunlop}, {Dwek}, {Eales}, {Farrah}, {F{\"o}rster Schreiber},
  {Fox}, {Franceschini}, {Gear}, {Genzel}, {Glenn}, {Griffin}, {Gruppioni},
  {Halpern}, {Hatziminaoglou}, {Ibar}, {Isaak}, {Ivison}, {Lagache}, {Le
  Borgne}, {Le Floc'h}, {Levenson}, {Lu}, {Lutz}, {Madden}, {Maffei}, {Magdis},
  {Mainetti}, {Maiolino}, {Marchetti}, {Mortier}, {Nguyen}, {Nordon},
  {O'Halloran}, {Okumura}, {Oliver}, {Omont}, {Page}, {Panuzzo},
  {Papageorgiou}, {Pearson}, {Perez Fournon}, {P{\'e}rez Garc{\'{\i}}a},
  {Poglitsch}, {Pohlen}, {Popesso}, {Pozzi}, {Rawlings}, {Rigopoulou},
  {Riguccini}, {Rizzo}, {Rodighiero}, {Roseboom}, {Rowan-Robinson},
  {Saintonge}, {Sanchez Portal}, {Santini}, {Sauvage}, {Schulz}, {Scott},
  {Seymour}, {Shao}, {Shupe}, {Smith}, {Stevens}, {Sturm}, {Symeonidis},
  {Tacconi}, {Trichas}, {Tugwell}, {Vaccari}, {Valtchanov}, {Vieira},
  {Vigroux}, {Wang}, {Ward}, {Wright}, {Xu}, \& {Zemcov}}]{elbaz:10}
{Elbaz}, D., {Hwang}, H.~S., {Magnelli}, B., {et~al.} 2010, \aap, 518, L29+

\bibitem[{{Erben} {et~al.}(2009){Erben}, {Hildebrandt}, {Lerchster}, {Hudelot},
  {Benjamin}, {van Waerbeke}, {Schrabback}, {Brimioulle}, {Cordes}, {Dietrich},
  {Holhjem}, {Schirmer}, \& {Schneider}}]{erben:09}
{Erben}, T., {Hildebrandt}, H., {Lerchster}, M., {et~al.} 2009, \aap, 493, 1197

\bibitem[{{Faber} {et~al.}(2007){Faber}, {Willmer}, {Wolf}, {Koo}, {Weiner},
  {Newman}, {Im}, {Coil}, {Conroy}, {Cooper}, {Davis}, {Finkbeiner}, {Gerke},
  {Gebhardt}, {Groth}, {Guhathakurta}, {Harker}, {Kaiser}, {Kassin},
  {Kleinheinrich}, {Konidaris}, {Kron}, {Lin}, {Luppino}, {Madgwick},
  {Meisenheimer}, {Noeske}, {Phillips}, {Sarajedini}, {Schiavon}, {Simard},
  {Szalay}, {Vogt}, \& {Yan}}]{faber:07}
{Faber}, S.~M., {Willmer}, C.~N.~A., {Wolf}, C., {et~al.} 2007, \apj, 665, 265

\bibitem[{{Fioc} \& {Rocca-Volmerange}(1997)}]{pegase}
{Fioc}, M., \& {Rocca-Volmerange}, B. 1997, \aap, 326, 950

\bibitem[{{Fontana} {et~al.}(2006){Fontana}, {Salimbeni}, {Grazian},
  {Giallongo}, {Pentericci}, {Nonino}, {Fontanot}, {Menci}, {Monaco},
  {Cristiani}, {Vanzella}, {de Santis}, \& {Gallozzi}}]{fontana:06}
{Fontana}, A., {Salimbeni}, S., {Grazian}, A., {et~al.} 2006, \aap, 459, 745

\bibitem[{{Franx} {et~al.}(2008){Franx}, {van Dokkum}, {Schreiber}, {Wuyts},
  {Labb{\'e}}, \& {Toft}}]{franx:08}
{Franx}, M., {van Dokkum}, P.~G., {Schreiber}, N.~M.~F., {et~al.} 2008, \apj,
  688, 770

\bibitem[{{Hildebrandt}(2010)}]{hildebrandt:10}
{Hildebrandt}, H. 2010, \aap, 99

\bibitem[{{Hildebrandt} {et~al.}(2009){Hildebrandt}, {Pielorz}, {Erben}, {van
  Waerbeke}, {Simon}, \& {Capak}}]{hildebrandt:09}
{Hildebrandt}, H., {Pielorz}, J., {Erben}, T., {et~al.} 2009, \aap, 498, 725

\bibitem[{{Hopkins} {et~al.}(2010){Hopkins}, {Bundy}, {Croton}, {Hernquist},
  {Keres}, {Khochfar}, {Stewart}, {Wetzel}, \& {Younger}}]{hopkins:10}
{Hopkins}, P.~F., {Bundy}, K., {Croton}, D., {et~al.} 2010, \apj, 715, 202

\bibitem[{{Ilbert} {et~al.}(2009){Ilbert}, {Capak}, {Salvato}, {Aussel},
  {McCracken}, {Sanders}, {Scoville}, {Kartaltepe}, {Arnouts}, {Floc'h},
  {Mobasher}, {Taniguchi}, {Lamareille}, {Leauthaud}, {Sasaki}, {Thompson},
  {Zamojski}, {Zamorani}, {Bardelli}, {Bolzonella}, {Bongiorno}, {Brusa},
  {Caputi}, {Carollo}, {Contini}, {Cook}, {Coppa}, {Cucciati}, {de la Torre},
  {de Ravel}, {Franzetti}, {Garilli}, {Hasinger}, {Iovino}, {Kampczyk},
  {Kneib}, {Knobel}, {Kovac}, {LeBorgne}, {LeBrun}, {F{\`e}vre}, {Lilly},
  {Looper}, {Maier}, {Mainieri}, {Mellier}, {Mignoli}, {Murayama}, {Pell{\`o}},
  {Peng}, {P{\'e}rez-Montero}, {Renzini}, {Ricciardelli}, {Schiminovich},
  {Scodeggio}, {Shioya}, {Silverman}, {Surace}, {Tanaka}, {Tasca}, {Tresse},
  {Vergani}, \& {Zucca}}]{ilbert:09}
{Ilbert}, O., {Capak}, P., {Salvato}, M., {et~al.} 2009, \apj, 690, 1236

\bibitem[{{Ilbert} {et~al.}(2010){Ilbert}, {Salvato}, {Le Floc'h}, {Aussel},
  {Capak}, {McCracken}, {Mobasher}, {Kartaltepe}, {Scoville}, {Sanders},
  {Arnouts}, {Bundy}, {Cassata}, {Kneib}, {Koekemoer}, {Le F{\`e}vre}, {Lilly},
  {Surace}, {Taniguchi}, {Tasca}, {Thompson}, {Tresse}, {Zamojski}, {Zamorani},
  \& {Zucca}}]{ilbert:10}
{Ilbert}, O., {Salvato}, M., {Le Floc'h}, E., {et~al.} 2010, \apj, 709, 644

\bibitem[{{Juneau} {et~al.}(2005){Juneau}, {Glazebrook}, {Crampton},
  {McCarthy}, {Savaglio}, {Abraham}, {Carlberg}, {Chen}, {Le Borgne}, {Marzke},
  {Roth}, {J{\"o}rgensen}, {Hook}, \& {Murowinski}}]{juneau:05}
{Juneau}, S., {Glazebrook}, K., {Crampton}, D., {et~al.} 2005, \apjl, 619, L135

\bibitem[{{Kauffmann} {et~al.}(2003){Kauffmann}, {Heckman}, {White}, {Charlot},
  {Tremonti}, {Peng}, {Seibert}, {Brinkmann}, {Nichol}, {SubbaRao}, \&
  {York}}]{kauffmann:03}
{Kauffmann}, G., {Heckman}, T.~M., {White}, S.~D.~M., {et~al.} 2003, \mnras,
  341, 54

\bibitem[{{Kennicutt}(1998)}]{kennicutt:98}
{Kennicutt}, Jr., R.~C. 1998, \araa, 36, 189

\bibitem[{{Kormendy} \& {Djorgovski}(1989)}]{kormendy:89}
{Kormendy}, J., \& {Djorgovski}, S. 1989, \araa, 27, 235

\bibitem[{{Kriek} {et~al.}(2010){Kriek}, {Labb{\'e}}, {Conroy}, {Whitaker},
  {van Dokkum}, {Brammer}, {Franx}, {Illingworth}, {Marchesini}, {Muzzin},
  {Quadri}, \& {Rudnick}}]{kriek:10}
{Kriek}, M., {Labb{\'e}}, I., {Conroy}, C., {et~al.} 2010, \apjl, 722, L64

\bibitem[{{Kriek} {et~al.}(2008){Kriek}, {van der Wel}, {van Dokkum}, {Franx},
  \& {Illingworth}}]{kriek:08}
{Kriek}, M., {van der Wel}, A., {van Dokkum}, P.~G., {et~al.} 2008, \apj, 682,
  896

\bibitem[{{Kriek} {et~al.}(2006){Kriek}, {van Dokkum}, {Franx}, {Quadri},
  {Gawiser}, {Herrera}, {Illingworth}, {Labb{\'e}}, {Lira}, {Marchesini},
  {Rix}, {Rudnick}, {Taylor}, {Toft}, {Urry}, \& {Wuyts}}]{kriek:06b}
{Kriek}, M., {van Dokkum}, P.~G., {Franx}, M., {et~al.} 2006, \apjl, 649, L71

\bibitem[{{Kriek} {et~al.}(2009){Kriek}, {van Dokkum}, {Labb{\'e}}, {Franx},
  {Illingworth}, {Marchesini}, \& {Quadri}}]{kriek:09a}
{Kriek}, M., {van Dokkum}, P.~G., {Labb{\'e}}, I., {et~al.} 2009, \apj, 700,
  221

\bibitem[{{Kriek} {et~al.}(2011){Kriek}, {van Dokkum}, {Whitaker}, {Labbe},
  {Franx}, \& {Brammer}}]{kriek:11}
{Kriek}, M., {van Dokkum}, P.~G., {Whitaker}, K.~E., {et~al.} 2011, ArXiv
  e-prints

\bibitem[{{Kroupa}(2001)}]{kroupa:01}
{Kroupa}, P. 2001, \mnras, 322, 231

\bibitem[{{Labb{\'e}} {et~al.}(2006){Labb{\'e}}, {Bouwens}, {Illingworth}, \&
  {Franx}}]{labbe:06}
{Labb{\'e}}, I., {Bouwens}, R., {Illingworth}, G.~D., {et~al.} 2006, \apjl,
  649, L67

\bibitem[{{Labb{\'e}} {et~al.}(2007){Labb{\'e}}, {Franx}, {Rudnick},
  {Schreiber}, {van Dokkum}, {Moorwood}, {Rix}, {R{\"o}ttgering}, {Trujillo},
  \& {van der Werf}}]{labbe:07}
{Labb{\'e}}, I., {Franx}, M., {Rudnick}, G., {et~al.} 2007, \apj, 665, 944

\bibitem[{{Labb{\'e}} {et~al.}(2005){Labb{\'e}}, {Huang}, {Franx}, {Rudnick},
  {Barmby}, {Daddi}, {van Dokkum}, {Fazio}, {Schreiber}, {Moorwood}, {Rix},
  {R{\"o}ttgering}, {Trujillo}, \& {van der Werf}}]{labbe:05}
{Labb{\'e}}, I., {Huang}, J., {Franx}, M., {et~al.} 2005, \apjl, 624, L81

\bibitem[{{Le Floc'h} {et~al.}(2009){Le Floc'h}, {Aussel}, {Ilbert},
  {Riguccini}, {Frayer}, {Salvato}, {Arnouts}, {Surace}, {Feruglio},
  {Rodighiero}, {Capak}, {Kartaltepe}, {Heinis}, {Sheth}, {Yan}, {McCracken},
  {Thompson}, {Sanders}, {Scoville}, \& {Koekemoer}}]{lefloch:09}
{Le Floc'h}, E., {Aussel}, H., {Ilbert}, O., {et~al.} 2009, \apj, 703, 222

\bibitem[{{Le Floc'h} {et~al.}(2005){Le Floc'h}, {Papovich}, {Dole}, {Bell},
  {Lagache}, {Rieke}, {Egami}, {P{\'e}rez-Gonz{\'a}lez}, {Alonso-Herrero},
  {Rieke}, {Blaylock}, {Engelbracht}, {Gordon}, {Hines}, {Misselt}, {Morrison},
  \& {Mould}}]{lefloch:05}
{Le Floc'h}, E., {Papovich}, C., {Dole}, H., {et~al.} 2005, \apj, 632, 169

\bibitem[{{Lilly} {et~al.}(2007){Lilly}, {Le F{\`e}vre}, {Renzini}, {Zamorani},
  {Scodeggio}, {Contini}, {Carollo}, {Hasinger}, {Kneib}, {Iovino}, {Le Brun},
  {Maier}, {Mainieri}, {Mignoli}, {Silverman}, {Tasca}, {Bolzonella},
  {Bongiorno}, {Bottini}, {Capak}, {Caputi}, {Cimatti}, {Cucciati}, {Daddi},
  {Feldmann}, {Franzetti}, {Garilli}, {Guzzo}, {Ilbert}, {Kampczyk}, {Kovac},
  {Lamareille}, {Leauthaud}, {Borgne}, {McCracken}, {Marinoni}, {Pello},
  {Ricciardelli}, {Scarlata}, {Vergani}, {Sanders}, {Schinnerer}, {Scoville},
  {Taniguchi}, {Arnouts}, {Aussel}, {Bardelli}, {Brusa}, {Cappi}, {Ciliegi},
  {Finoguenov}, {Foucaud}, {Franceschini}, {Halliday}, {Impey}, {Knobel},
  {Koekemoer}, {Kurk}, {Maccagni}, {Maddox}, {Marano}, {Marconi}, {Meneux},
  {Mobasher}, {Moreau}, {Peacock}, {Porciani}, {Pozzetti}, {Scaramella},
  {Schiminovich}, {Shopbell}, {Smail}, {Thompson}, {Tresse}, {Vettolani},
  {Zanichelli}, \& {Zucca}}]{lilly:07}
{Lilly}, S.~J., {Le F{\`e}vre}, O., {Renzini}, A., {et~al.} 2007, \apjs, 172,
  70

\bibitem[{{Ma{\'{\i}}z Apell{\'a}niz}(2006)}]{maiz:06}
{Ma{\'{\i}}z Apell{\'a}niz}, J. 2006, \aj, 131, 1184

\bibitem[{{Maller} {et~al.}(2009){Maller}, {Berlind}, {Blanton}, \&
  {Hogg}}]{maller:09}
{Maller}, A.~H., {Berlind}, A.~A., {Blanton}, M.~R., {et~al.} 2009, \apj, 691,
  394

\bibitem[{{Maraston}(2005)}]{maraston:05}
{Maraston}, C. 2005, \mnras, 362, 799

\bibitem[{{Maraston} {et~al.}(2010){Maraston}, {Pforr}, {Renzini}, {Daddi},
  {Dickinson}, {Cimatti}, \& {Tonini}}]{maraston:10}
{Maraston}, C., {Pforr}, J., {Renzini}, A., {et~al.} 2010, \mnras, 407, 830

\bibitem[{{Marchesini} {et~al.}(2009){Marchesini}, {van Dokkum}, {F{\"o}rster
  Schreiber}, {Franx}, {Labb{\'e}}, \& {Wuyts}}]{marchesini:09}
{Marchesini}, D., {van Dokkum}, P.~G., {F{\"o}rster Schreiber}, N.~M., {et~al.}
  2009, \apj, 701, 1765

\bibitem[{{Marchesini} {et~al.}(2010){Marchesini}, {Whitaker}, {Brammer}, {van
  Dokkum}, \& {Labb\'e}}]{marchesini:10}
{Marchesini}, D., {Whitaker}, K.~E., {Brammer}, G.~B., {et~al.} 2010, \apj

\bibitem[{{Martin} {et~al.}(2007){Martin}, {Wyder}, {Schiminovich}, {Barlow},
  {Forster}, {Friedman}, {Morrissey}, {Neff}, {Seibert}, {Small}, {Welsh},
  {Bianchi}, {Donas}, {Heckman}, {Lee}, {Madore}, {Milliard}, {Rich}, {Szalay},
  \& {Yi}}]{martin:07}
{Martin}, D.~C., {Wyder}, T.~K., {Schiminovich}, D., {et~al.} 2007, \apjs, 173,
  342

\bibitem[{{Muzzin} {et~al.}(2009){Muzzin}, {Marchesini}, {van Dokkum},
  {Labb{\'e}}, {Kriek}, \& {Franx}}]{muzzin:09}
{Muzzin}, A., {Marchesini}, D., {van Dokkum}, P.~G., {et~al.} 2009, \apj, 701,
  1839

\bibitem[{{Muzzin} {et~al.}(2010){Muzzin}, {van Dokkum}, {Kriek}, {Labb{\'e}},
  {Cury}, {Marchesini}, \& {Franx}}]{muzzin:10}
{Muzzin}, A., {van Dokkum}, P., {Kriek}, M., {et~al.} 2010, \apj, 725, 742

\bibitem[{{Papovich} {et~al.}(2011){Papovich}, {Finkelstein}, {Ferguson},
  {Lotz}, \& {Giavalisco}}]{papovich:11}
{Papovich}, C., {Finkelstein}, S.~L., {Ferguson}, H.~C., {et~al.} 2011, \mnras,
  412, 1123

\bibitem[{{Papovich} {et~al.}(2006){Papovich}, {Moustakas}, {Dickinson}, {Le
  Floc'h}, {Rieke}, {Daddi}, {Alexander}, {Bauer}, {Brandt}, {Dahlen}, {Egami},
  {Eisenhardt}, {Elbaz}, {Ferguson}, {Giavalisco}, {Lucas}, {Mobasher},
  {P{\'e}rez-Gonz{\'a}lez}, {Stutz}, {Rieke}, \& {Yan}}]{papovich:06}
{Papovich}, C., {Moustakas}, L.~A., {Dickinson}, M., {et~al.} 2006, \apj, 640,
  92

\bibitem[{{Papovich} {et~al.}(2007){Papovich}, {Rudnick}, {Le Floc'h}, {van
  Dokkum}, {Rieke}, {Taylor}, {Armus}, {Gawiser}, {Huang}, {Marcillac}, \&
  {Franx}}]{papovich:07}
{Papovich}, C., {Rudnick}, G., {Le Floc'h}, E., {et~al.} 2007, \apj, 668, 45

\bibitem[{{Pozzetti} {et~al.}(2010){Pozzetti}, {Bolzonella}, {Zucca},
  {Zamorani}, {Lilly}, {Renzini}, {Moresco}, {Mignoli}, {Cassata}, {Tasca},
  {Lamareille}, {Maier}, {Meneux}, {Halliday}, {Oesch}, {Vergani}, {Caputi},
  {Kova{\v c}}, {Cimatti}, {Cucciati}, {Iovino}, {Peng}, {Carollo}, {Contini},
  {Kneib}, {Le F{\'e}vre}, {Mainieri}, {Scodeggio}, {Bardelli}, {Bongiorno},
  {Coppa}, {de la Torre}, {de Ravel}, {Franzetti}, {Garilli}, {Kampczyk},
  {Knobel}, {Le Borgne}, {Le Brun}, {Pell{\`o}}, {Perez Montero},
  {Ricciardelli}, {Silverman}, {Tanaka}, {Tresse}, {Abbas}, {Bottini}, {Cappi},
  {Guzzo}, {Koekemoer}, {Leauthaud}, {Maccagni}, {Marinoni}, {McCracken},
  {Memeo}, {Porciani}, {Scaramella}, {Scarlata}, \& {Scoville}}]{pozzetti:10}
{Pozzetti}, L., {Bolzonella}, M., {Zucca}, E., {et~al.} 2010, \aap, 523, A13+

\bibitem[{{Probst} {et~al.}(2004){Probst}, {Gaughan}, {Abraham}, {Andrew},
  {Daly}, {Hileman}, {Hunten}, {Liang}, {Merrill}, {Repp}, \&
  {Shaw}}]{probst:newfirm}
{Probst}, R.~G., {Gaughan}, N., {Abraham}, M., {et~al.} 2004, in Presented at
  the Society of Photo-Optical Instrumentation Engineers (SPIE) Conference,
  Vol. 5492, Society of Photo-Optical Instrumentation Engineers (SPIE)
  Conference Series, ed. {A.~F.~M.~Moorwood \& M.~Iye}, 1716--1724

\bibitem[{{Quadri} {et~al.}(2007){Quadri}, {van Dokkum}, {Gawiser}, {Franx},
  {Marchesini}, {Lira}, {Rudnick}, {Herrera}, {Maza}, {Kriek}, {Labb{\'e}}, \&
  {Francke}}]{quadri:07}
{Quadri}, R., {van Dokkum}, P., {Gawiser}, E., {et~al.} 2007, \apj, 654, 138

\bibitem[{{Rudnick} {et~al.}(2003){Rudnick}, {Rix}, {Franx}, {Labb{\'e}},
  {Blanton}, {Daddi}, {F{\"o}rster Schreiber}, {Moorwood}, {R{\"o}ttgering},
  {Trujillo}, {van de Wel}, {van der Werf}, {van Dokkum}, \& {van
  Starkenburg}}]{rudnick:03}
{Rudnick}, G., {Rix}, H.-W., {Franx}, M., {et~al.} 2003, \apj, 599, 847

\bibitem[{{Rudnick} {et~al.}(2009){Rudnick}, {von der Linden}, {Pell{\'o}},
  {Arag{\'o}n-Salamanca}, {Marchesini}, {Clowe}, {De Lucia}, {Halliday},
  {Jablonka}, {Milvang-Jensen}, {Poggianti}, {Saglia}, {Simard}, {White}, \&
  {Zaritsky}}]{rudnick:09}
{Rudnick}, G., {von der Linden}, A., {Pell{\'o}}, R., {et~al.} 2009, \apj, 700,
  1559

\bibitem[{{Salpeter}(1955)}]{salpeter:55}
{Salpeter}, E.~E. 1955, \apj, 121, 161

\bibitem[{{Sanders} {et~al.}(2007){Sanders}, {Salvato}, {Aussel}, {Ilbert},
  {Scoville}, {Surace}, {Frayer}, {Sheth}, {Helou}, {Brooke}, {Bhattacharya},
  {Yan}, {Kartaltepe}, {Barnes}, {Blain}, {Calzetti}, {Capak}, {Carilli},
  {Carollo}, {Comastri}, {Daddi}, {Ellis}, {Elvis}, {Fall}, {Franceschini},
  {Giavalisco}, {Hasinger}, {Impey}, {Koekemoer}, {Le F{\`e}vre}, {Lilly},
  {Liu}, {McCracken}, {Mobasher}, {Renzini}, {Rich}, {Schinnerer}, {Shopbell},
  {Taniguchi}, {Thompson}, {Urry}, \& {Williams}}]{sanders:07}
{Sanders}, D.~B., {Salvato}, M., {Aussel}, H., {et~al.} 2007, \apjs, 172, 86

\bibitem[{{Schechter}(1976)}]{schechter:76}
{Schechter}, P. 1976, \apj, 203, 297

\bibitem[{{Scoville} {et~al.}(2007){Scoville}, {Aussel}, {Brusa}, {Capak},
  {Carollo}, {Elvis}, {Giavalisco}, {Guzzo}, {Hasinger}, {Impey}, {Kneib},
  {LeFevre}, {Lilly}, {Mobasher}, {Renzini}, {Rich}, {Sanders}, {Schinnerer},
  {Schminovich}, {Shopbell}, {Taniguchi}, \& {Tyson}}]{scoville:cosmos}
{Scoville}, N., {Aussel}, H., {Brusa}, M., {et~al.} 2007, \apjs, 172, 1

\bibitem[{{Somerville} {et~al.}(2004){Somerville}, {Lee}, {Ferguson},
  {Gardner}, {Moustakas}, \& {Giavalisco}}]{somerville:04b}
{Somerville}, R.~S., {Lee}, K., {Ferguson}, H.~C., {et~al.} 2004, \apjl, 600,
  L171

\bibitem[{{Strateva} {et~al.}(2001){Strateva}, {Ivezi{\'c}}, {Knapp},
  {Narayanan}, {Strauss}, {Gunn}, {Lupton}, {Schlegel}, {Bahcall}, {Brinkmann},
  {Brunner}, {Budav{\'a}ri}, {Csabai}, {Castander}, {Doi}, {Fukugita}, {Gy{\H
  o}ry}, {Hamabe}, {Hennessy}, {Ichikawa}, {Kunszt}, {Lamb}, {McKay},
  {Okamura}, {Racusin}, {Sekiguchi}, {Schneider}, {Shimasaku}, \&
  {York}}]{strateva:01}
{Strateva}, I., {Ivezi{\'c}}, {\v Z}., {Knapp}, G.~R., {et~al.} 2001, \aj, 122,
  1861

\bibitem[{{Taylor} {et~al.}(2009{\natexlab{a}}){Taylor}, {Franx}, {van Dokkum},
  {Bell}, {Brammer}, {Rudnick}, {Wuyts}, {Gawiser}, {Lira}, {Urry}, \&
  {Rix}}]{taylor:09a}
{Taylor}, E.~N., {Franx}, M., {van Dokkum}, P.~G., {et~al.} 2009{\natexlab{a}},
  \apj, 694, 1171

\bibitem[{{Taylor} {et~al.}(2009{\natexlab{b}}){Taylor}, {Franx}, {van Dokkum},
  {Quadri}, {Gawiser}, {Bell}, {Barrientos}, {Blanc}, {Castander}, {Damen},
  {Gonzalez-Perez}, {Hall}, {Herrera}, {Hildebrandt}, {Kriek}, {Labb{\'e}},
  {Lira}, {Maza}, {Rudnick}, {Treister}, {Urry}, {Willis}, \&
  {Wuyts}}]{taylor:09b}
---. 2009{\natexlab{b}}, \apjs, 183, 295

\bibitem[{{van Dokkum}(2005)}]{vandokkum:05}
{van Dokkum}, P.~G. 2005, \aj, 130, 2647

\bibitem[{{van Dokkum} {et~al.}(2009){van Dokkum}, {Labb{\'e}}, {Marchesini},
  {Quadri}, {Brammer}, {Whitaker}, {Kriek}, {Franx}, {Rudnick}, {Illingworth},
  {Lee}, \& {Muzzin}}]{nmbs}
{van Dokkum}, P.~G., {Labb{\'e}}, I., {Marchesini}, D., {et~al.} 2009, \pasp,
  121, 2

\bibitem[{{van Dokkum} {et~al.}(2010){van Dokkum}, {Whitaker}, {Brammer},
  {Franx}, {Kriek}, {Labb{\'e}}, {Marchesini}, {Quadri}, {Bezanson},
  {Illingworth}, {Muzzin}, {Rudnick}, {Tal}, \& {Wake}}]{vandokkum:10}
{van Dokkum}, P.~G., {Whitaker}, K.~E., {Brammer}, G., {et~al.} 2010, \apj,
  709, 1018

\bibitem[{{Wake} {et~al.}(2006){Wake}, {Nichol}, {Eisenstein}, {Loveday},
  {Edge}, {Cannon}, {Smail}, {Schneider}, {Scranton}, {Carson}, {Ross},
  {Brunner}, {Colless}, {Couch}, {Croom}, {Driver}, {da {\^A}ngela}, {Jester},
  {de Propris}, {Drinkwater}, {Bland-Hawthorn}, {Pimbblet}, {Roseboom},
  {Shanks}, {Sharp}, \& {Brinkmann}}]{wake:06}
{Wake}, D.~A., {Nichol}, R.~C., {Eisenstein}, D.~J., {et~al.} 2006, \mnras,
  372, 537

\bibitem[{{Whitaker} {et~al.}(2011){Whitaker}, {van Dokkum}, {Brammer},
  {Kriek}, {Franx}, {Labb\'e}, {Quadri}, {Bezanson}, {Illingworth}, {Lee},
  {Muzzin}, {Rudnick}, \& {Wake}}]{whitaker:11}
{Whitaker}, K., {van Dokkum}, P., {Brammer}, G., {et~al.} 2011, \apj, submitted

\bibitem[{{Whitaker} {et~al.}(2010){Whitaker}, {van Dokkum}, {Brammer},
  {Kriek}, {Franx}, {Labb{\'e}}, {Marchesini}, {Quadri}, {Bezanson},
  {Illingworth}, {Lee}, {Muzzin}, {Rudnick}, \& {Wake}}]{whitaker:10}
{Whitaker}, K.~E., {van Dokkum}, P.~G., {Brammer}, G., {et~al.} 2010, \apj,
  719, 1715

\bibitem[{{Williams} {et~al.}(2009){Williams}, {Quadri}, {Franx}, {van Dokkum},
  \& {Labb{\'e}}}]{williams:09}
{Williams}, R.~J., {Quadri}, R.~F., {Franx}, M., {et~al.} 2009, \apj, 691, 1879

\bibitem[{{Wolf} {et~al.}(2003){Wolf}, {Meisenheimer}, {Rix}, {Borch}, {Dye},
  \& {Kleinheinrich}}]{wolf:03}
{Wolf}, C., {Meisenheimer}, K., {Rix}, H.-W., {et~al.} 2003, \aap, 401, 73

\bibitem[{{Wuyts} {et~al.}(2007){Wuyts}, {Labb{\'e}}, {Franx}, {Rudnick}, {van
  Dokkum}, {Fazio}, {F{\"o}rster Schreiber}, {Huang}, {Moorwood}, {Rix},
  {R{\"o}ttgering}, \& {van der Werf}}]{wuyts:07}
{Wuyts}, S., {Labb{\'e}}, I., {Franx}, M., {et~al.} 2007, \apj, 655, 51

\bibitem[{{Wuyts} {et~al.}(2008){Wuyts}, {Labb{\'e}}, {Schreiber}, {Franx},
  {Rudnick}, {Brammer}, \& {van Dokkum}}]{wuyts:fireworks}
{Wuyts}, S., {Labb{\'e}}, I., {Schreiber}, N.~M.~F., {et~al.} 2008, \apj, 682,
  985

\bibitem[{{Wyder} {et~al.}(2007){Wyder}, {Martin}, {Schiminovich}, {Seibert},
  {Budav{\'a}ri}, {Treyer}, {Barlow}, {Forster}, {Friedman}, {Morrissey},
  {Neff}, {Small}, {Bianchi}, {Donas}, {Heckman}, {Lee}, {Madore}, {Milliard},
  {Rich}, {Szalay}, {Welsh}, \& {Yi}}]{wyder:07}
{Wyder}, T.~K., {Martin}, D.~C., {Schiminovich}, D., {et~al.} 2007, \apjs, 173,
  293

\end{thebibliography}

%%%%%%%%%%%%%%%%%%%%%%%%%%%%%%%%%%%%%%%%%%%%%%%%%%%%%%%%%%%%%%%%%%
%
%   Appendices
%
%%%%%%%%%%%%%%%%%%%%%%%%%%%%%%%%%%%%%%%%%%%%%%%%%%%%%%%%%%%%%%%%%%
\begin{appendix}

\section{Line emission in the broad/medium-band filters}
\label{ap:emlines}

%%%%%%%%%%%%%%%%%%%%%% Emission line contamination     %%%%%%%%%%%%%%%%%%%%
\begin{figure}
	\epsscale{0.7}
	\plotone{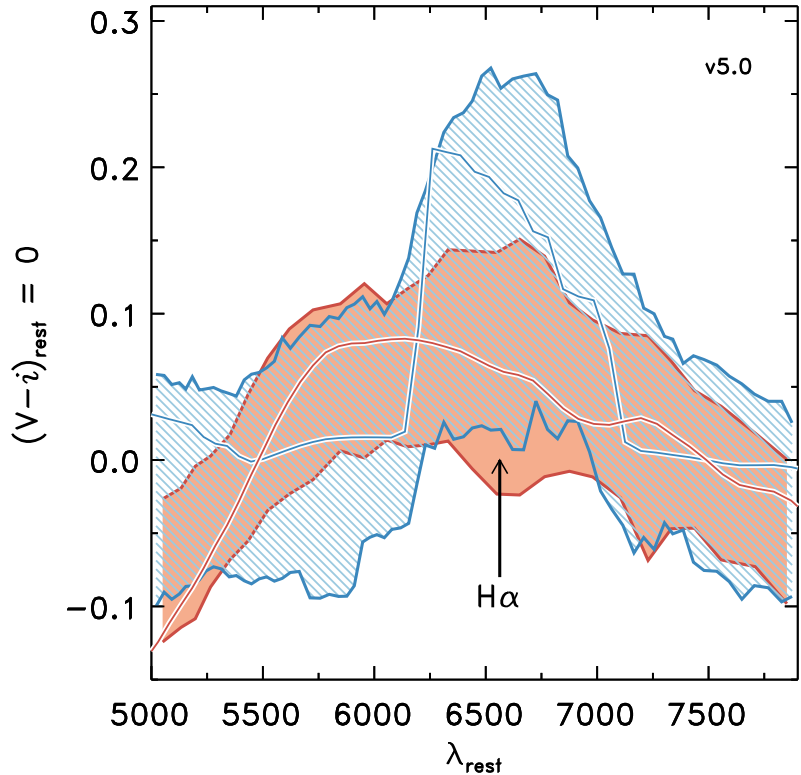}
	\caption[Emission line contamination of broad/medium-band fluxes]{Emission line contamination of broad/medium-band fluxes. The broad-band photometry of each object with $0.4<z<2.2$ and $\log M/M_\odot > 9.5$ are shifted into the rest-frame and the SEDs are normalized by subtracting a linear fit to the computed rest-frame $V$ and $i$-band fluxes.  Star-forming (blue) and quiescent (red) galaxies are selected as in Fig. \ref{f:color_color}, and the shaded areas show the 1-$\sigma$ range of the observed fluxes shifted into the rest-frame.  An $H\alpha$ emission feature is clearly visible for the galaxies selected to be star-forming, even though this feature is only crudely sampled by the $i\mbox{--}K$ broad- and medium-band filters over the redshift range shown.   The thin red and blue lines show two \eazy\ templates integrated through the NEWFIRM $J_2$ filter at $0 < z <4$.  \label{f:emission_lines}}
\end{figure}

We find that adjusting the treatment of nebular emission lines in the
\eazy\ templates, following \cite{ilbert:10}, significantly improves
the quality of our photometric redshift estimates.  This treatment
becomes even more important as we use medium-band photometry from
4000\AA\ to 1.7\micron\ and the medium-width filters are more
sensitive to the flux contamination by emission lines.  We investigate
the contribution of emission lines to the medium- and broad-band
photometry in Fig. \ref{f:emission_lines}.  The shaded regions show
the distributions of broad-band fluxes shifted to the rest-frame of
objects in the quiescent and star-forming populations, selected as in
Fig.\ref{f:red_sequence} (see \cite{whitaker:10} for a more detailed
description of the construction of these ``rest-frame SEDs''.)  We
include galaxies over the full redshift range of the NMBS and $\log
M/M_\odot > 9.5$.  We detect the clear signature of H$\alpha$
emission, with a shape similar to that expected from a typical filter
sensitivity curve, in the average SED of the star-forming sample.  This
feature is sampled in the figure by, among others, the $J_1$, $J_3$,
and $H_2$ filters at $z=0.6, 0.9,~\mathrm{and}~1.6$, respectively.
The quiescent population shows no such feature and has a different overall
SED shape from the star-forming population.  These results highlight both the
importance of including emission lines in the redshift templates and
also that our star-forming/quiescent
selection criterion succeeds in selecting galaxies
with/without H$\alpha$ emission. The detection of spectral
features in the NMBS photometry, including the shape and strength of the H$\alpha$ emission line as a function of galaxy SED type, is explored in greater detail by \citet{kriek:11}.

\section{Effect of mass measurement errors on the density evolution}

\begin{figure*}
	\plotone{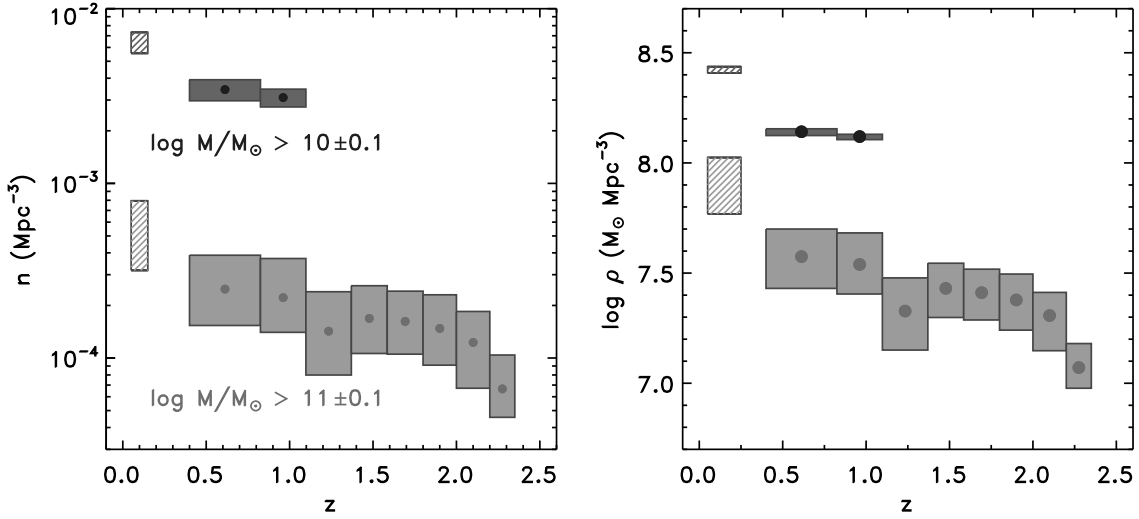}
	\caption[Effect of mass measurement errors on number and mass density estimates]{Effect of mass measurement errors on number and mass density estimates.  The shaded areas indicate the range of densities measured adopting the indicated mass cuts plus or minus a systematic measurement ``error'' of 0.1 dex.\label{f:mass_errors}}
\end{figure*}

Redshift-dependent systematic errors in the stellar masses are an
important systematic uncertainty in our study. It is very difficult to
assess how important such errors are. As noted in, e.g.,
\citet{marchesini:09} the metallicities, IMF, and other parameters
could vary systematically with redshift. Furthermore, we sample a different
part of ``model space'' at low redshift (where we rely on 
models for 10+ Gyr old stellar populations) and high redshift
(where we rely on models for younger populations, and star-forming
galaxies). In Fig.\ \ref{f:mass_errors} we show the effects of
a 0.1 dex error in the masses on the number densities and the
mass densities. The effects are substantial, particularly for
the number densities at the highest masses.
A 0.1 dex redshift-dependent error implies a range in
the evolution of the number density of massive galaxies from
very little to a decrease of
a factor of $\sim$5 between $z=0.4$ and $z=2$.
The only way to
reduce these uncertainties is to obtain dynamical mass measurements
of large samples of high redshift galaxies.

%%%%%%%%%%%%%%%%%  Effect of mass measurement errors on density   %%%%%%  
%%%%%%%%%%%%%%%%%%%%%%%%%%%%%%%%%%%%%%%%%%%%%%%%%%%%%%%%%%%%%%%%%%%%%%

\begin{deluxetable}{lccll}
\tabletypesize{\footnotesize}
\tablecolumns{5} 
\tablewidth{0pt} 
\tablecaption{Number and mass densities}
\tablehead{\colhead{}                                    &
           \multicolumn{2}{c}{$\log M/M_\odot > 10$}        &
           \multicolumn{2}{c}{$\log M/M_\odot > 11$}        \\
		   \colhead{$z$}          &
		   \colhead{$n$\tablenotemark{a}}          &
		   \colhead{$\rho$\tablenotemark{a}}       &
		   \colhead{$n$}          &
		   \colhead{$\rho$}   }
\startdata
%\sidehead{Large Magellanic Cloud }
0.61 & 34.4$\pm$5.2 & 13.9$\pm$2.1 & 2.5$\pm$0.6   	 & 3.8$\pm$0.9 \\
0.96 & 31.0$\pm$4.7 & 13.2$\pm$2.0 & 2.2$\pm$0.6   	 & 3.5$\pm$0.9 \\
1.23 & $\cdots$     &  $\cdots$    & 1.4$\pm$0.3   	 & 2.1$\pm$0.5 \\
1.48 & $\cdots$     &  $\cdots$    & 1.7$\pm$0.4   	 & 2.7$\pm$0.6 \\
1.69 & $\cdots$     &  $\cdots$	   & 1.6$\pm$0.4   	 & 2.6$\pm$0.6 \\
1.90 & $\cdots$     &  $\cdots$	   & 1.5$\pm$0.3   	 & 2.4$\pm$0.6 \\
2.10 & $\cdots$     &  $\cdots$    & 1.2$\pm$0.3   	 & 2.0$\pm$0.5 \\
2.28 & $\cdots$     &  $\cdots$    & 0.7$\pm$0.2   	 & 1.2$\pm$0.3
\enddata

\tablenotetext{a}{$n$: $10^{-4}\ \mathrm{Mpc}^{-3}$, $\rho$: $10^7\ M_\odot\ \mathrm{Mpc}^{-3}$}
\label{t:tab1}
\end{deluxetable}

\begin{deluxetable}{lccccccccllll}
\tabletypesize{\footnotesize}
\tablecolumns{13} 
\tablewidth{0pt} 
\tablecaption{Number and mass densities, split by galaxy type}
\tablehead{\colhead{}                                    &
	       \multicolumn{4}{c}{$10.2 < \log M/M_\odot < 10.6$}     &
           \multicolumn{4}{c}{$10.6 < \log M/M_\odot < 11.0$}     &
           \multicolumn{4}{c}{$11.0 < \log M/M_\odot < 11.6$}     \\
		   \colhead{}          &
		   \multicolumn{2}{c}{$n$\tablenotemark{a}} & 
		   \multicolumn{2}{c}{$\rho$\tablenotemark{a}} & 
		   \multicolumn{2}{c}{$n$} & 
		   \multicolumn{2}{c}{$\rho$} & 
		   \multicolumn{2}{c}{$n$} & 
		   \multicolumn{2}{c}{$\rho$} \\
		   \colhead{$z$}          &
		   \colhead{Q\tablenotemark{b}}       &
		   \colhead{SF\tablenotemark{b}}      &
		   \colhead{Q}       &
		   \colhead{SF}      &
		   \colhead{Q}       &
		   \colhead{SF}      &
		   \colhead{Q}       &
		   \colhead{SF}      &
		   \colhead{Q}       &
		   \colhead{SF}      &
		   \colhead{Q}       &
		   \colhead{SF}      }
\startdata
0.613 &    7.2$\pm$   1.4 &    7.3$\pm$   1.4 &    1.9$\pm$   0.4 &    1.8$\pm$   0.3 &      	 6.5$\pm$   1.3 &    2.1$\pm$   0.5 &    4.1$\pm$   0.8 &    1.2$\pm$   0.3 &   	  2.2$\pm$   0.5 &    0.3$\pm$   0.1 &    3.4$\pm$   0.8 &    0.3$\pm$   0.2 \\   
0.961 &    6.0$\pm$   1.2 &    6.3$\pm$   1.3 &    1.6$\pm$   0.3 &    1.6$\pm$   0.3 &      	 5.9$\pm$   1.2 &    3.1$\pm$   0.7 &    3.8$\pm$   0.8 &    1.8$\pm$   0.4 &   	  1.9$\pm$   0.5 &    0.3$\pm$   0.1 &    3.0$\pm$   0.8 &    0.4$\pm$   0.2 \\   
1.234 &    3.1$\pm$   0.6 &    6.3$\pm$   1.0 &    0.8$\pm$   0.2 &    1.6$\pm$   0.3 &      	 3.1$\pm$   0.6 &    2.1$\pm$   0.4 &    2.0$\pm$   0.4 &    1.3$\pm$   0.3 &   	  1.0$\pm$   0.3 &    0.4$\pm$   0.1 &    1.5$\pm$   0.4 &    0.5$\pm$   0.2 \\  
1.477 &     $\cdots$      &     $\cdots$      &     $\cdots$      &     $\cdots$      &      	 2.9$\pm$   0.5 &    2.4$\pm$   0.5 &    1.8$\pm$   0.3 &    1.4$\pm$   0.3 &   	  1.2$\pm$   0.3 &    0.5$\pm$   0.1 &    2.0$\pm$   0.5 &    0.7$\pm$   0.2 \\   
1.692 &     $\cdots$      &     $\cdots$      &     $\cdots$      &     $\cdots$      &      	 2.4$\pm$   0.5 &    2.7$\pm$   0.5 &    1.5$\pm$   0.3 &    1.6$\pm$   0.3 &   	  1.0$\pm$   0.3 &    0.6$\pm$   0.2 &    1.7$\pm$   0.4 &    0.8$\pm$   0.3 \\   
1.900 &     $\cdots$      &     $\cdots$      &     $\cdots$      &     $\cdots$      &      	  $\cdots$      &     $\cdots$      &     $\cdots$      &     $\cdots$      &   	  0.7$\pm$   0.2 &    0.8$\pm$   0.2 &    1.1$\pm$   0.3 &    1.2$\pm$   0.3 \\   
2.100 &     $\cdots$      &     $\cdots$      &     $\cdots$      &     $\cdots$      &      	  $\cdots$      &     $\cdots$      &     $\cdots$      &     $\cdots$      &   	  0.6$\pm$   0.2 &    0.6$\pm$   0.2 &    0.9$\pm$   0.3 &    0.9$\pm$   0.3 \\   
2.275 &     $\cdots$      &     $\cdots$      &     $\cdots$      &     $\cdots$      &      	  $\cdots$      &     $\cdots$      &     $\cdots$      &     $\cdots$      &   	  0.2$\pm$   0.1 &    0.4$\pm$   0.1 &    0.3$\pm$   0.1 &    0.7$\pm$   0.2 \\
2.425 &     $\cdots$      &     $\cdots$      &     $\cdots$      &     $\cdots$      &   	  $\cdots$      &     $\cdots$      &     $\cdots$      &     $\cdots$      &   	  0.2$\pm$   0.1 &    0.5$\pm$   0.2 &    0.3$\pm$   0.1 &    0.9$\pm$   0.3    
\enddata

\tablenotetext{a}{$n$: $10^{-4}\ \mathrm{Mpc}^{-3}$, $\rho$: $10^7\ M_\odot\ \mathrm{Mpc}^{-3}$}
\tablenotetext{b}{Q: quiescent, SF: star-forming (Fig. \ref{f:color_color}).}
\label{t:tab2}
\end{deluxetable}

\end{appendix}

\end{document}